\documentclass[lettersize,journal]{IEEEtran}
\newcommand{\papertitle}{The Fused Kernel Library: A C++ API to Develop Highly-Efficient GPU Libraries}
\usepackage{amsmath,amsfonts}
\usepackage[T1]{fontenc}
\usepackage{siunitx}

\usepackage{algorithmicx}
\usepackage{algorithm}
\usepackage{array}
\usepackage[scaled=0.85]{beramono}
\usepackage[font=small,labelfont=bf]{caption}
\usepackage[shortlabels]{enumitem}
\usepackage{subcaption}
\usepackage{textcomp}
\usepackage{stfloats}
\usepackage{url}
\usepackage{verbatim}
\usepackage{graphicx}
\usepackage{tabularx}
\usepackage[acronym]{glossaries}
\usepackage[automake]{glossaries-extra}
\usepackage[pdftex,
            pdfauthor={Oscar Amoros, Albert Andaluz, Johnny Nuñez and Antonio J. Peña},
            pdftitle={\papertitle},
            pdfsubject={Library paper},
            pdfkeywords={CUDA, Kernel Fusion, Vertical Fusion, Horizontal Fusion},
            pdfproducer={Latex with hyperref},
            pdfcreator={pdflatex},
            hidelinks]{hyperref}
\usepackage{xcolor}
\usepackage{algorithmicx}
\usepackage{algpseudocode}

\usepackage[USenglish]{babel}
\usepackage{numprint}
\usepackage{csquotes}
\usepackage[auth-lg]{authblk}
\definecolor{LightGray}{gray}{0.9}
\usepackage{orcidlink}
\usepackage[backend=biber,sorting=none,style=ieee]{biblatex}
\graphicspath{ {images/} }
\makeglossaries
\setabbreviationstyle[acronym]{long}
\npthousandsep{,}

\begin{document}	
	\title{\papertitle}	

\author{Oscar Amoros\thanks{O. Amoros is with Departament d’Arquitectura de Computadors, Universitat Politecnica de
Catalunya, Barcelona, Spain (email: oscar.amoros.huguet@upc.edu)},
Albert Andaluz\thanks{A. Andaluz is an independent researcher, Barcelona, Spain (email: albertandaluz@hotmail.com)},
Johnny Nuñez\thanks{J. Nuñez is with NVIDIA Computing SL, Madrid, Spain (email: johnunez@nvidia.com)},
Antonio J. Peña
\IEEEmembership{Senior, IEEE}\thanks{A. J. Peña is with the Barcelona Supecomputing Center (BSC) (email: antonio.pena@bsc.es)}
\thanks{This work has been submitted to the IEEE for possible publication. Copyright may be transferred without notice, after which this version may no longer be accessible.}
}
	\maketitle
        \newacronym{alu}{ALU}{Arithmetic Logic Unit}
        \newacronym{api}{API}{Application Programming Interface}
	\newacronym{kf}{KF}{Kernel Fusion}
        \newacronym{fk}{FK}{Fused Kernel}
	\newacronym{hf}{HF}{Horizontal Fusion}
	\newacronym{vf}{VF}{Vertical Fusion} 
	\newacronym{op}{Op}{Operation}
	\newacronym{iop}{IOp}{Instantiable Operation}
	\newacronym{uiop}{UIOp}{Unary Instantiable Operation}
	\newacronym{biop}{BIOp}{Binary Instantiable Operation}
	\newacronym{cop}{COp}{Compute Operation}
	\newacronym{mop}{MOp}{Memory Operation}  
	\newacronym{uop}{UOp}{Unary Operation}  
	\newacronym{bop}{BOp}{Binary Operation}  
	\newacronym{rop}{ROp}{Read Operation}  
	\newacronym{wop}{WOp}{Write Operation} 
	\newacronym{mbk}{MBK}{Memory Bound Kernel}
	\newacronym{cbk}{CBK}{Compute Bound Kernel}
        \newacronym{mb}{MB}{Memory Bound}
	\newacronym{cb}{CB}{Compute Bound}
	\newacronym{dpp}{DPP}{Data Parallel Pattern}
	\newacronym{riop}{RIOp}{Read Instantiable Operation}
	\newacronym{wiop}{WIOp}{Write Instantiable Operation}
    \newacronym{cc}{CC}{Compute Capability} 
    \newacronym{sm}{SM}{Streaming Multiprocessor} 
    \newacronym{tb}{TB}{Thread Block}
    \newacronym{tc}{TC}{Thread Coarsening}    
    \newacronym{flopb}{FLOP/B}{Floating Point Operations per Byte read}
    \newacronym{npp}{NPP}{Nvidia Performance Primitives}
     \newacronym{fkl}{FKL}{Fused Kernel Library}
     \newacronym{cvgs}{cvGS}{cvGPUSpeedup}
     \newacronym{gpu}{GPU}{Graphics Processing Unit}
    \newacronym{cuda}{CUDA}{Compute Unified Device Architecture} 
    \newacronym{cpu}{CPU}{Central Processing Unit}
    \newacronym[shortplural={TIDs},longplural=Thread Indexes]{tid}{TIdx}{Thread Index}
  \newacronym{mu}{MU}{Methodology User}
  \newacronym{lu}{LU}{Library User}
	\begin{abstract}    
		 Existing GPU libraries often struggle to fully exploit the parallel resources and on-chip memory (SRAM) of \glsxtrshortpl{gpu} when chaining multiple GPU functions as individual kernels. While \glsxtrfull{kf} techniques like \glsxtrfull{hf} and \glsxtrfull{vf} can mitigate this, current library implementations often require library developers to manually create \glsxtrfullpl{fk}. Hence, library users rely on limited sets of pre-compiled or template-based fused kernels. This limits the use cases that can benefit from \glsxtrshort{hf}  and \glsxtrshort{vf} and increases development costs. In order to solve these issues, we present a novel methodology for building GPU libraries that enables automatic on-demand \glsxtrshort{hf} and \glsxtrshort{vf} for arbitrary combinations of GPU library functions. Our methodology defines reusable, fusionable components that users combine via high-level programming interfaces. Leveraging C++17 metaprogramming features available in compilers like nvcc, our methodology generates a single and optimized \glsxtrshort{fk} tailored to the user's specific sequence of operations at compile time, without needing a custom compiler or manual development and pre-compilation of kernel combinations. This approach abstracts low-level GPU complexities while maximizing GPU resource utilization and keeping intermediate data in SRAM. We provide an open-source implementation demonstrating significant speedups compared to traditional libraries in various benchmarks, validating the effectiveness of this methodology for improving GPU performance in the range of $2\times $ to more than $\numprint{1000} \times $, while preserving high-level programmability.
	\end{abstract}
	
    
	\section{Introduction}
	\IEEEPARstart{I}{n} the field of data-parallel computing, \glsxtrfullpl{gpu} have been dominating the architecture market growth for the last 20 years. Their rapid evolution, driven by economies of scale from the visualization and gaming markets, coupled with flexible programming \glsxtrfullpl{api} \cite{cuda2025,5457293,sycl2025,hip2025}, has made of this architecture the preferred choice for an increasing number of use cases, from clusters of massive GPU accelerators to workstations, PCs, laptops, and tiny edge devices.
    
	One of the primary reasons for this success is that \glsxtrshortpl{gpu} may perform data-parallel computations faster than \glsxtrshortpl{cpu}, while still supporting general-purpose programming. This superior parallel performance is attained by reducing cache sizes and control unit complexity, to allocate additional chip area for further \glsxtrfullpl{alu}, larger register banks, and a more programmable memory hierarchy \cite{pumps4th}. These changes require a programming model that is exposing many low-level complexities to the programmer. Hence, the number of software developers available that may efficiently program a \glsxtrshortpl{gpu} is limited.
    
	To soften this issue, GPU vendors and other companies have been developing a GPU library ecosystem that provides non-GPU programmers access to the vast performance capabilities of the GPU.	
	Noteworthy, none of these libraries solves the following issues in GPU library development at the same time:
	\begin{enumerate}
		\item{Fully exploit the parallel resources of the GPU.}
		\item{Fully exploit the GPU SRAM, keeping intermediate results in SRAM as possible.}
		\item{Do 1) and 2) for any combination of the library GPU functions, for any domain with a high-level \glsxtrshort{api}. }
        \item{Do 1), 2), and 3) without requiring a specialized compiler---only a standard C++ implementation. }
        \item{Do 1), 2), and 3) even when combining code from different libraries. }
	\end{enumerate}
	
    Many hand-crafted approaches \cite{DBLP:journals/cgf/EscaleraPAS11, DaoFERR22} apply a technique called \glsxtrfull{kf} \cite{wang2010}, which features two variants, \glsxtrfull{hf} \cite{9741270} and \glsxtrfull{vf} \cite{Wahib2014ScalableKF}.
    \glsxtrshort{vf} exploits temporal data locality among consecutive dependent data-parallel operations, by keeping the intermediate results in register banks, and \glsxtrshort{hf} maximizes the utilization of the data-parallel hardware resources, by parallelizing calls to the same kernel on independent and different data. The main drawback of the \glsxtrshort{vf} approach, is that it requires to manually implement \glsxtrshortpl{fk} as combinations of different kernels. In the case of \glsxtrshort{hf}, we encounter a similar situation: libraries implement manually-crafted versions of the fused kernels. This often requires impractical or unfeasible development efforts. Hence, production libraries tend to offer fused kernels only for the most frequent use cases. 
	Some state--of--the--art libraries attempt to apply these techniques~\cite{cvcuda2025,npp2025,tensorrt2025}; more advanced approaches use C++ templates to enable some kernel parametrization~\cite{cutlass2025,amdcomposable2022}. Still, these only offer a limited set of \glsxtrshortpl{fk} to the library user.
	
	By failing to address the aforementioned issues in a complete and generic manner, current libraries are either sub-optimal or too complex to be used by a broad audience. This creates a market for specialized GPU programmers and companies, that write hand-made GPU kernels for specific applications or provide alternative compilers that are costly to maintain.
    
	In this article, we present the building grounds of a novel methodology to write GPU libraries that addresses all the aforementioned issues. Our methodology provides a solution for automatic \glsxtrshort{hf} and \glsxtrshort{vf} in a generic approach, exposing fusionable and reusable components to the user that may be combined to create any type of kernel from a high-level API. As a consequence, the library creator is not required to write all kernel combinations. We attain this without the need for developing a new compiler, which would require to be maintained and updated with each new GPU architecture and language feature. Instead, our solution only relies on C++17 meta-programming features available in nvcc and any other standard-compliant compiler.
	
	Furthermore, we provide an open-source implementation, with a limited set of features, that enables us to imitate the syntax of any high-level GPU library, while providing important speedups. This implementation is already being used in production, creating optimized kernels by combining open-source with closed-source fusionable components that follow our methodology. Leveraging this open-source library, we explore the limits of the speedups attainable with the two types of \glsxtrshort{kf}, to better understand how fusion behaves according to different variables.
	
	Our results, attaining over $\numprint{20000}\times$ speedup in synthetic benchmarks, and $\numprint{200}\times$ speedup in production applications, with respect to OpenCV-CUDA and \glsxtrshort{npp} libraries, while mimicking their \glsxtrshortpl{api}, demonstrate that this methodology is effective in both \glsxtrshort{kf} and GPU library programmability. Moreover, our methodology may be expanded to tackle more complex algorithms in future work.
	
	\section{Background}
	
	GPUs represent a major shift in general-purpose architectures. Apart from their general-purpose capabilities that generated the popular GPGPU term, these are also specialized architectures for data-parallel applications. GPUs may perform task parallelism, although in a rather limited fashion and considerably slower than CPUs. Hence, GPUs are used as accelerators attached to host CPUs.
	
	Some of the key elements of the GPU architectures that enables them to feature massive parallel processing capabilities are:
	\begin{itemize}
		\item{Reduced cache memories.}
		\item{A complex programmable memory hierarchy \cite{pumps4th}.}
		\item{Markedly simple control units.}
	\end{itemize}
	
	These design choices enable to add more \glsxtrshortpl{alu}, providing vastly superior peak performance numbers with respect to their CPU counterparts. However, this poses a major impact on GPU DRAM latency and programmability. The reduced cache size and increased computational capacity exacerbate memory latency issues, leading to a more significant challenge for \glsxtrshortpl{gpu} than \glsxtrshortpl{cpu}. The programmable memory hierarchy and threading model also pose a significant challenge when trying to reach broad software development audiences. 

\begin{figure}
        \centering
        \includegraphics[keepaspectratio=true,width=0.48\textwidth]{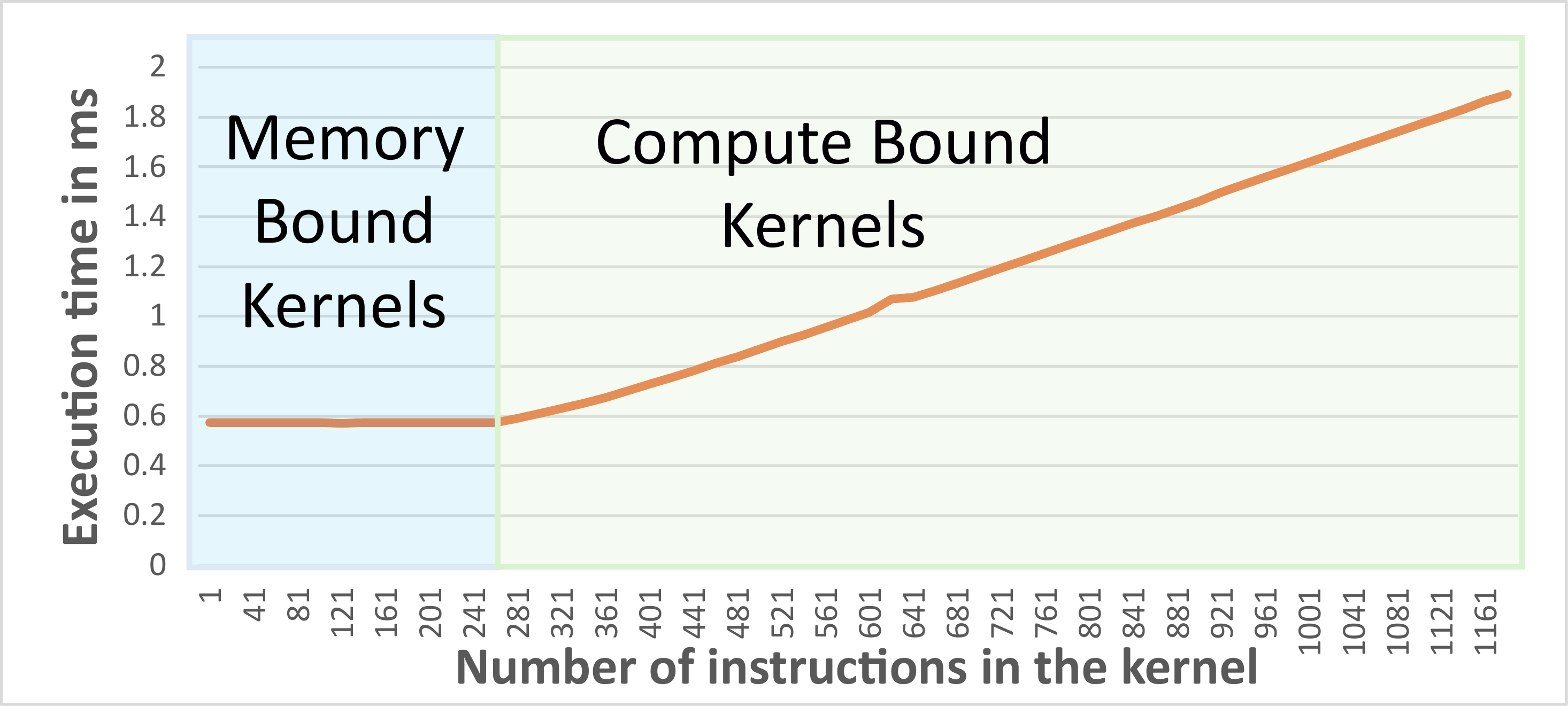}		
        \caption{Kernel execution time in milliseconds in an  RTX 4090 NVIDIA GPU, for a kernel with increasing number of instructions. The kernel is uni-dimensional, processing $N$ float elements where $N = 3840\times 2160\times 8$. This amount of data ensures that the memory bandwidth is consistently saturated. The instructions execute a simple addition ($\texttt{float} + \texttt{float}$) per thread, performed as many times as indicated in the {\tt x} axis of this figure (from 1 to 1,161). } 	
        \label{fig:LatencyHiding}
    \end{figure}

	To mitigate the latency problem, \glsxtrshortpl{gpu} incorporate a well-known technique known as latency hiding \cite{Volkov2016UnderstandingLH}. Latency hiding enables the GPU to execute memory reads and writes in parallel with arithmetical, logical, and other instructions, effectively hiding the time spent on read and write operations. As we can see in Figure~\ref{fig:LatencyHiding}, the overlapping only occurs when there are sufficient instructions per thread. Until that point, the factor determining the execution time of the kernels, is the time spent reading and writing data. Those kernels are classified as \glsxtrfull{mb}. As seen in the figure, when leveraging more than 260 instructions, the kernel execution time starts to increase every time we add more instructions. At that point, the kernels stop being \glsxtrshort{mb}, because the execution time now depends on the time to execute the 260+ instructions. Kernels with that condition are categorized as \glsxtrfull{cb}.
	
    Another mechanism \glsxtrshortpl{gpu} implement to try to cope with the increased computing capacity versus DRAM bandwidth, is by providing a DRAM memory system featuring wide aggregated bandwidth. Unfortunately, in order to fully utilize the GPU bandwidth, there needs to be sufficient GPU threads executing in parallel and reading in a coalesced fashion, a feature further defined by each GPU architecture.
	
	To further soften the bandwidth limitation, modern GPU architectures have increased the types of memory access patterns that lead to coalesced accesses, allowed different kernels to execute concurrently in the same GPU, and included some levels of transparent cache memory. But these optimizations are still insufficient for many use cases.
	
	In this context, it is apparent that in order to benefit from the \glsxtrshortpl{gpu} parallel compute power, it is necessary to establish a parallel programming model (such as CUDA~\cite{cuda2025}, OpenCL~\cite{5457293}, or SyCL~\cite{sycl2025}) that grants the programmer control over hardware behavior (mostly latencies and resource utilization). Providing this control requires the programmer to be aware of certain GPU hardware details, which renders GPU programming as a much more complex task than programming \glsxtrshortpl{cpu}. As a natural consequence, there shall be fewer software developers available with GPU programming knowledge.

    Due to the GPU programmability complexity, vast numbers of GPU libraries have been developed. These GPU libraries abstract the internals of GPU kernel programming away from users, in order to increase the number of programmers that may use \glsxtrshortpl{gpu} in their applications. 
	
	Despite many efforts, most GPU libraries do not properly use latency hiding and do not properly exploit the GPU parallel resources (aggregate DRAM bandwidth and parallel compute). In many cases, these issues are easy to be solved by the average CUDA programmer, handcrafting a custom CUDA kernel, usually refererd as \glsxtrfull{fk}. Unfortunately, even the average CUDA developer is a scarce resource.

\begin{figure}
		\centering
		\includegraphics[keepaspectratio=true,width=0.48\textwidth]{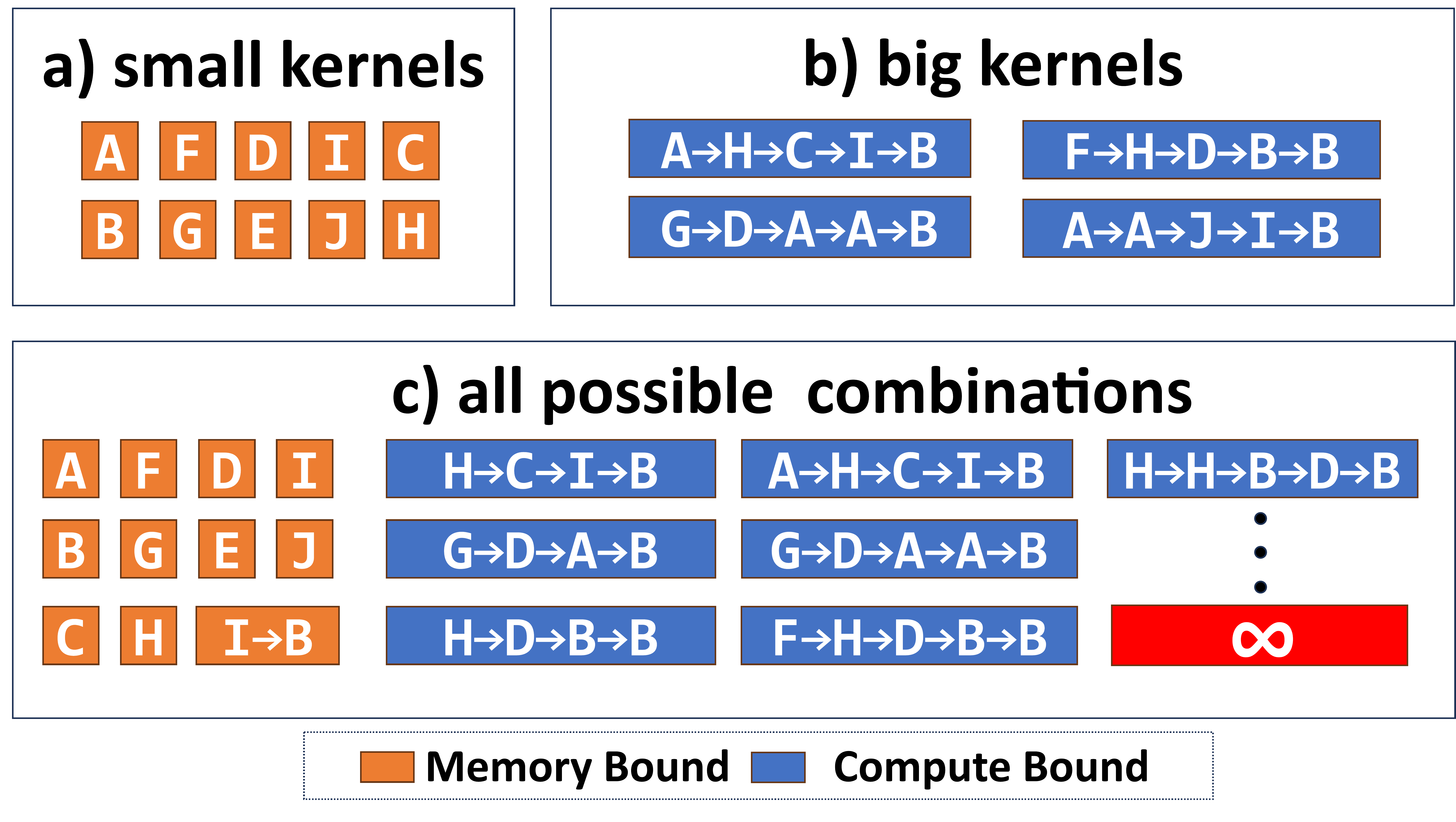}
		
		\caption{ Abstract representation of library functionality organization. Letters A to J represent different basic operations. Orange and blue boxes represent \glsxtrshort{mb} kernels and \glsxtrshort{cb} kernels, respectively. a) Represents libraries that offer many small \glsxtrshort{mb} kernels, which provides the library with user flexibility but poor performance. b) Represents libraries that contain more optimized code, but their users might not find the combination of operations they need. c) Represents all combinations of small and big kernels, while it is worth noting that featuring an infinite number of kernel implementations is unfeasible. } 	
		\label{fig:SlowUnflexibleExpensive}
	\end{figure}

\begin{figure}
		\centering		
        \includegraphics[keepaspectratio=true,width=0.48\textwidth]{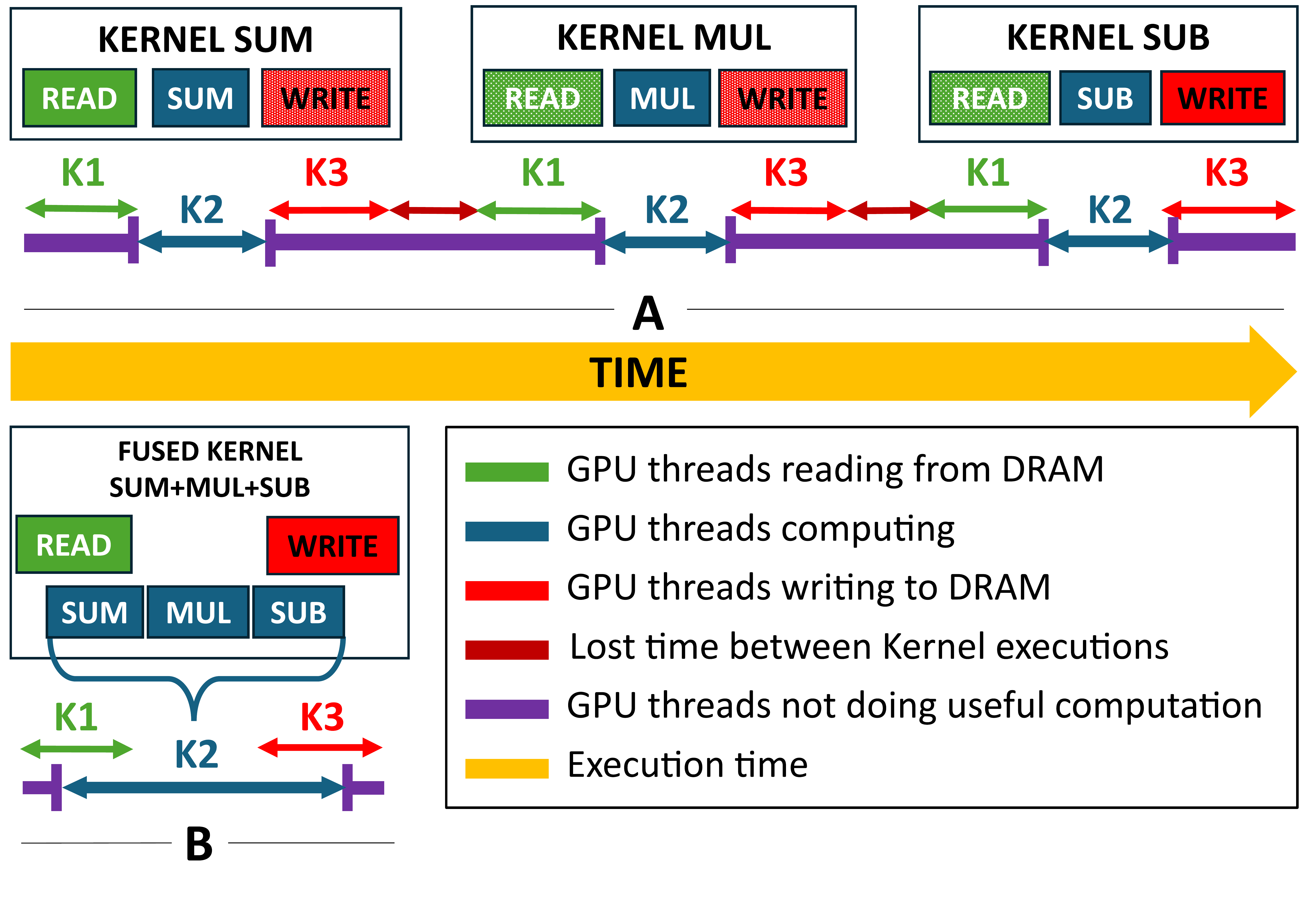}
		\caption{\glsxtrshort{vf} example. $K1$ represents the kernel code responsible for reading data from GPU DRAM. $K2$ represents the kernel code responsible for transforming the read data. $K3$ represents the kernel code responsible for storing the results on GPU DRAM. A) Launching 3 consecutive kernels SUM, MUL, and SUB. B) Launching a single Vertically Fused kernel $SUM+MUL+SUB$, that performs all the operations of the three kernels in A. It results in 1 DRAM read step ($K1$) and 1 DRAM write step ($K3$) instead of 3 reads and 3 writes, making the overall execution 3 times faster. This is possible because intermediate results are kept in GPU SRAM (in this case registers). We also show the overlapping of $K1$ and $K3$ with $K2$, thanks to latency hiding.}
		\label{fig:verticalfusion}
	\end{figure}

	Libraries of type a) (small kernels) seen in Figure~\ref{fig:SlowUnflexibleExpensive} implement large numbers of \glsxtrshort{mb} kernels. When those kernels are executed one after the other with data dependencies~\cite{CompArch2019}, as seen in Figure~\ref{fig:verticalfusion}A and Algorithm \ref{alg:kernelcall} lines 1 to 8, the result of each kernel will not stay in the GPU SRAM; it will be written to off-chip DRAM. As we can see in Figure~\ref{fig:verticalfusion}B and Algorithm \ref{alg:kernelcall} lines from 9 to 16, if instead of deploying separate kernels we implement the same operations in a single \glsxtrshort{fk}, the resulting execution time is much shorter because the intermediate results of SUM and MUL stay on registers (SRAM) instead, which prevents a worst case latency of 300 to 600 GPU computing cycles \cite{Volkov2016UnderstandingLH} per DRAM memory access, depending on the GPU architecture and model.

      \begin{algorithm}
       \caption{Example of inefficient code (MainFunctionA and KernelFunctionA) versus its efficient counterpart   (MainFunctionB and KernelFunctionB).}
        \label{alg:kernelcall}
		\begin{algorithmic}[1]			\Function{KernelFunctionA}{}
			\State \Call{Operation}{}
			\EndFunction
            
			\Function{MainFunctionA}{}
			\For{$i \gets 1$ to $N$}
            
			\State \Call{KernelFunctionA}{}
			\EndFor
			\EndFunction			
			\Function{KernelFunctionB}{}
			\For{$i \gets 1$ to $N$}
			\State \Call{Operation}{}
			\EndFor 
			\EndFunction			
			\Function{MainFunctionB}{}
			\State \Call{KernelFunctionB}{}
			\EndFunction			
		\end{algorithmic}     
	\end{algorithm}
    
        Additionally, the resulting \glsxtrshort{fk} will perform further operations between the first read and the final write operation. This will improve the utilization of the GPU's latency hiding, by executing the additional operations from the subsequent kernels at markedly low relative cost due to the absence of memory operations, until the \glsxtrshort{fk} becomes \glsxtrshort{cb}. Even when the \glsxtrshort{fk} becomes \glsxtrshort{cb}, the execution time increase when adding one instruction to the \glsxtrshort{fk} is in the range of a few microseconds, while adding the same instruction in a consecutive separate kernel would add 600 microseconds, as seen in Figure~\ref{fig:LatencyHiding}. The optimization Figure~\ref{fig:verticalfusion}B and Algorithm \ref{alg:kernelcall} lines from 9 to 16 represent, is usually referred as \glsxtrshort{vf}.

\begin{figure}
        \centering
            \begin{subfigure}{0.23\textwidth}
                \includegraphics[keepaspectratio=true,width=1.0\textwidth]{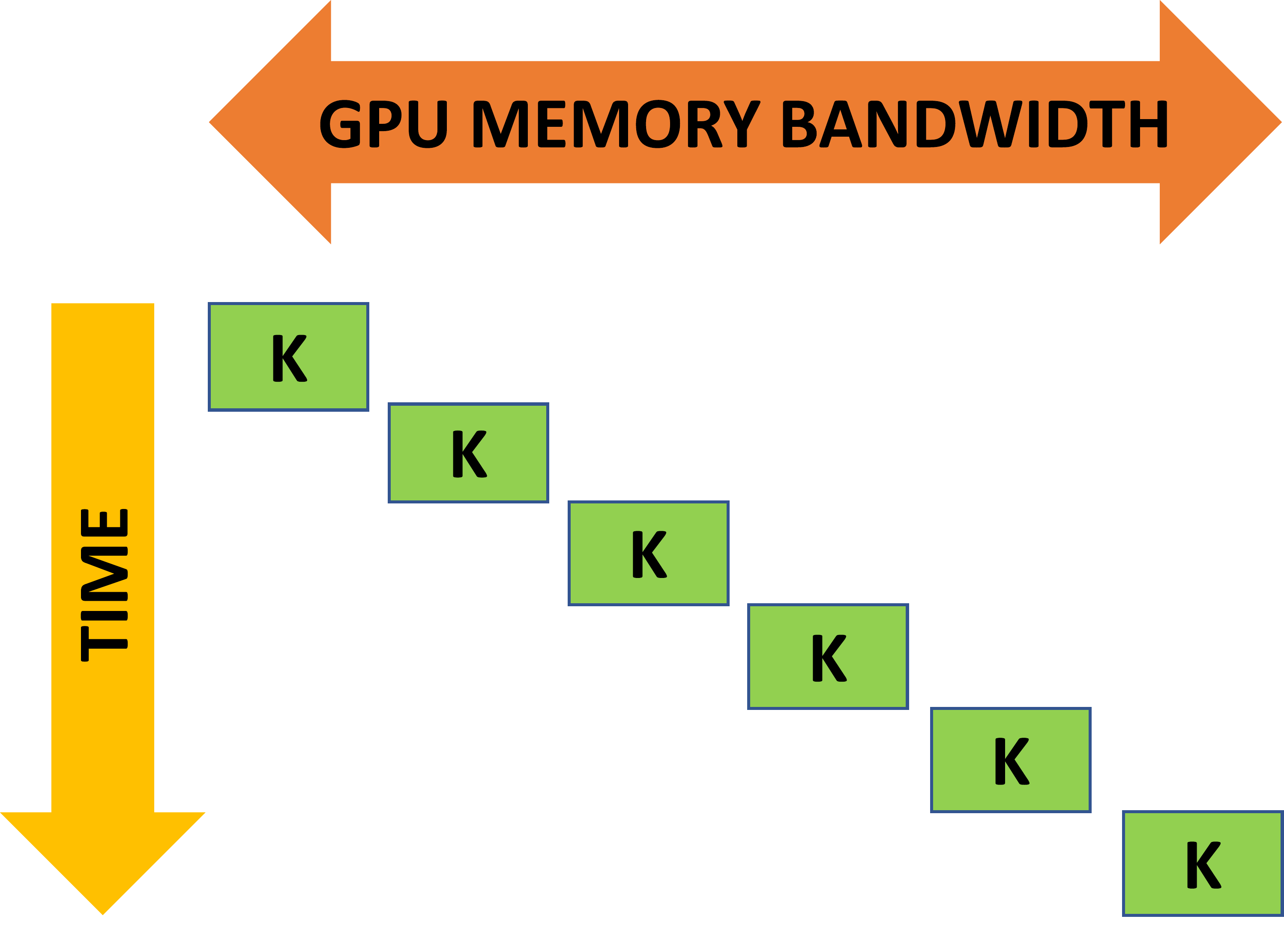}
                \caption{Multiple kernels}		
                \label{fig:horizontalmultikernel}
            \end{subfigure}
            \begin{subfigure}{0.24\textwidth}
                \includegraphics[keepaspectratio=true,width=1.0\textwidth]{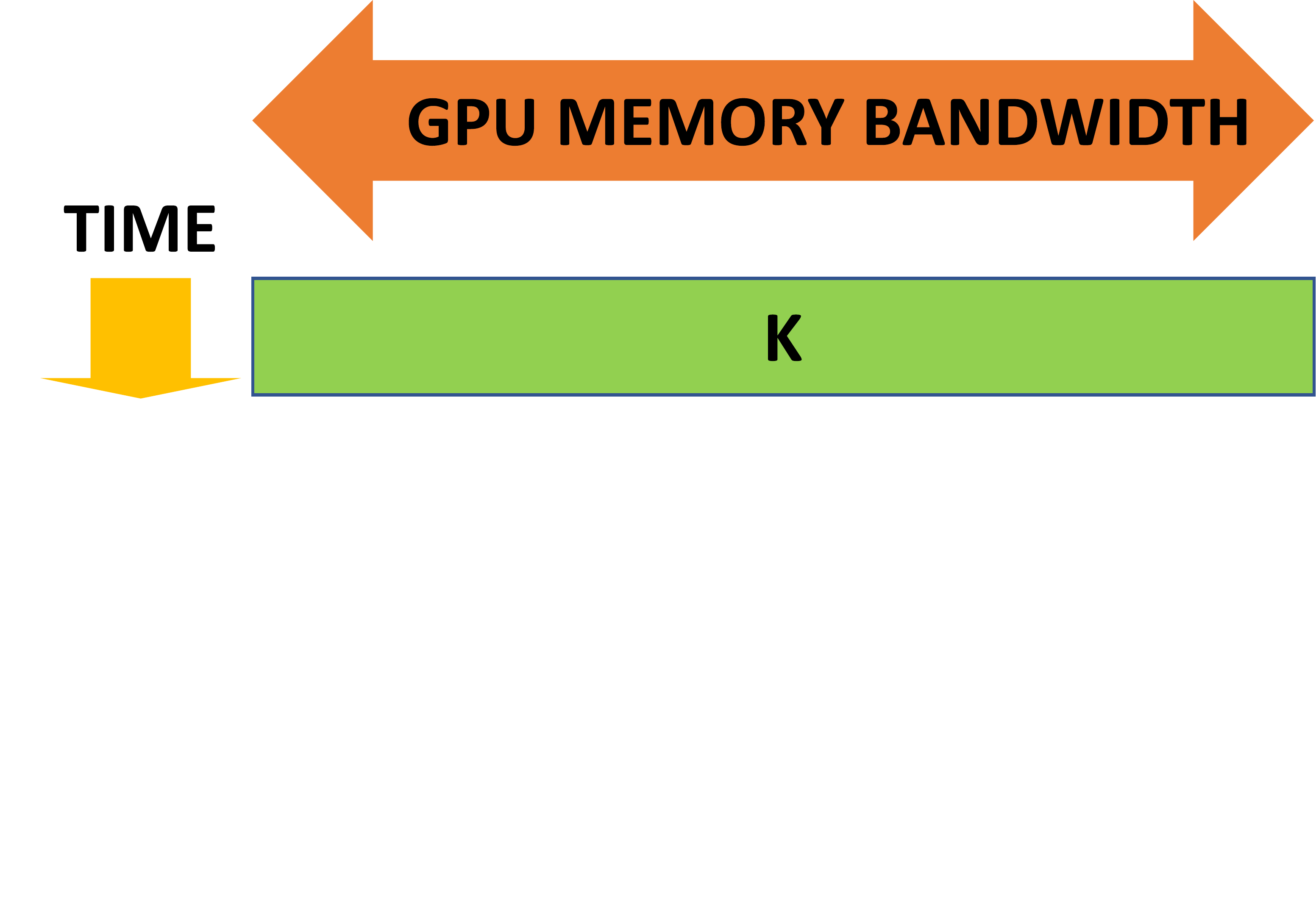}
                \caption{Single kernel}
                \label{fig:horizontalsinglekernel}
            \end{subfigure}
        \caption{\glsxtrshort{hf} example. Figure~\ref{fig:horizontalmultikernel}: Several independent calls to the same kernel on different data, that execute sequentially. Figure~\ref{fig:horizontalsinglekernel}: A single kernel executing the same code for 5 planes of threads, each reading from different data, resulting in concurrent memory accesses and leading to an improved utilization of the GPU DRAM memory bandwidth, resulting in a considerably reduced execution time.}
        \end{figure}

        Besides improving latency hiding utilization in GPU libraries, often the performance issue these libraries experience is related to the GPU resource utilization. The most common case is that where the user of the library needs to launch many independent calls of the same kernel on different data, with a limited number of GPU threads on each call. These kernels will be executed sequentially and will use a small portion of the GPU's total resources (DRAM bandwidth and compute), as seen in Figure~\ref{fig:horizontalmultikernel}, limiting overall performance.
    
        \label{fig:horizontalfusionexample}

\begin{figure}
		\centering
		\includegraphics[width=0.35\textwidth]{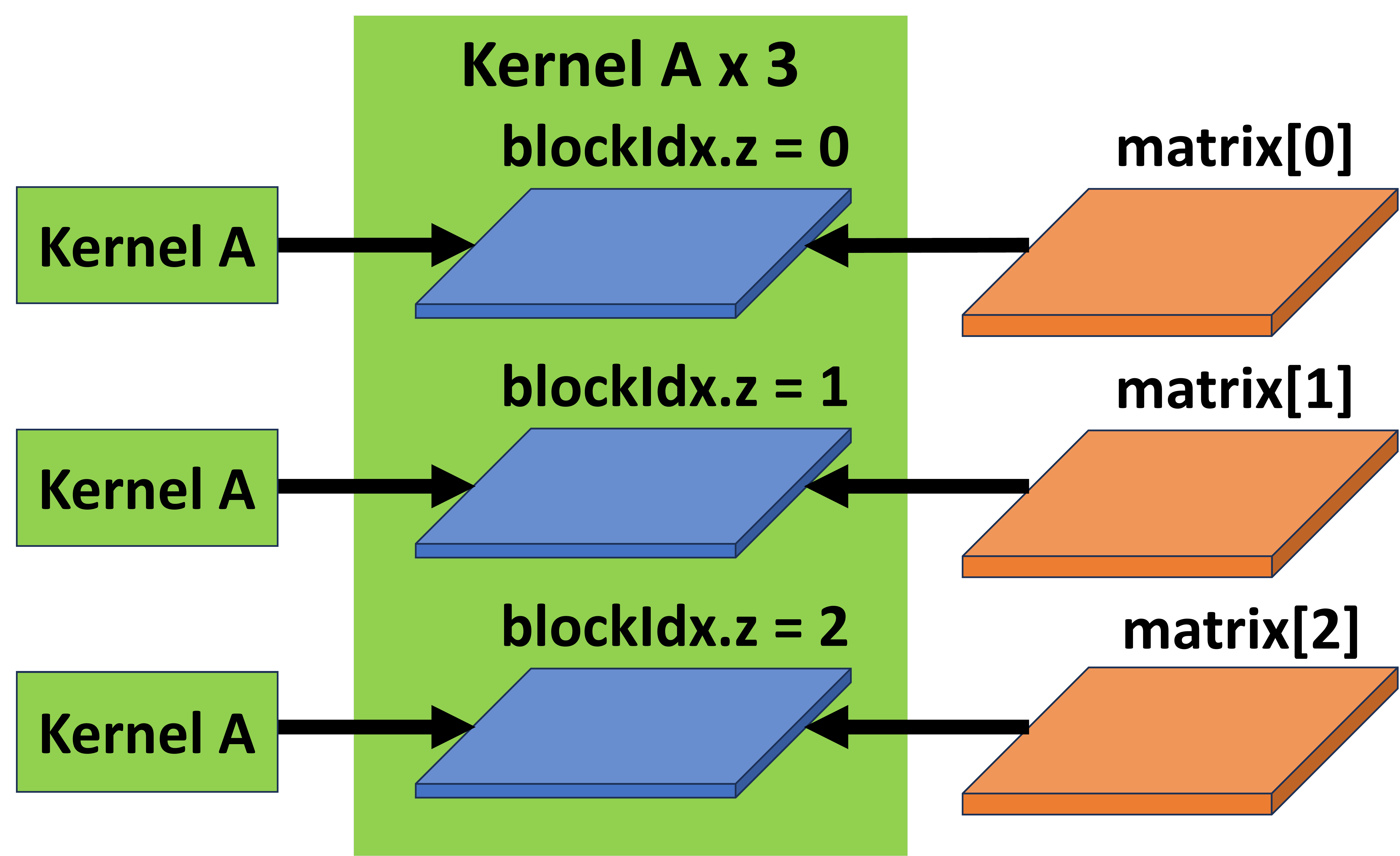}
		\caption{ Representation of \glsxtrshort{hf} for a kernel named A, that will be executing 3 times in parallel, each time on a different data matrix. We are representing kernel A 3 times, to imply that despite featuring a single horizontally fused kernel, each thread plane of that kernel will be executing exactly the same code. The value in \texttt{blockIdx.z}, will be used by each plane to determine the region of memory it needs to read.
		}
		\label{fig:horizontalfusion}
	\end{figure}

        Consequently, some libraries adopt \glsxtrshort{hf} \cite{npp2025, cvcuda2025}. \glsxtrshort{hf} solves or mitigates the GPU resource utilization concern, often using the third dimension of the kernel grid to determine which data needs to use each \glsxtrfull{tb} plane, as seen in Figure~\ref{fig:horizontalfusion}, effectively transforming each of the kernels into one of the \glsxtrshort{tb} planes of the \glsxtrshort{fk}, as seen in Figure~\ref{fig:horizontalsinglekernel} and Figure~\ref{fig:horizontalfusion}. 

    \section{Related work}
        Traditionally GPU libraries implement \glsxtrshort{vf} and \glsxtrshort{hf} following methodologies that can be grouped in three categories.
    
     \paragraph{Pre-compiled \glsxtrshort{vf} and \glsxtrshort{hf}} Providing some vertically and or horizontally fused kernels, written by hand by the library creators \cite{cvcuda2025,npp2025, flashinfer2025} as in Figure~\ref{fig:SlowUnflexibleExpensive}a or a small sub set of Figure~\ref{fig:SlowUnflexibleExpensive}c. This solution only provides improved performance for a small set of use cases and greatly increases development and maintainability costs.
    
    \paragraph{Composable Kernels} Converting some specific kernels into configurable or composable \cite{amdcomposable2022}, so that it is possible to add point-wise operations at the beginning and\slash or at the end of the kernel to perform \glsxtrshort{vf} and the possibility to configure \glsxtrshort{hf}. This approach does not define a general mechanism for this type of fusion that may be used to create kernels. It is domain-specific unlike our methodology. Moreover, it does not solve the problem of combining GPU code in a single kernel, from different libraries, while using a high-level API.
        
    \paragraph{Specialized Compilers} Alternative solutions to implement both \glsxtrshort{vf} and \glsxtrshort{hf} are based on the creation of a GPU compiler, specialized in automatically performing that fusion \cite{9741270, 10.1145/2082156.2082183, verticalfusion_10.1145/3571284, FractalTensor, modular2025mojo, verticalfusion_10.1145/3571284}. However, these compiler-based solutions require even more specialized knowledge and very costly maintenance, and they are sometimes domain-specific. Additionally, these increase the problem of code duplication, in this case at the compiler level, and do not address code fusion from different libraries.

        An alternative to implement \glsxtrshort{hf} would be to use a different stream for each kernel. However, kernel execution may well be faster than stream command insertion in CPU. In this case, CUDA Graphs \cite{NVIDIA_CUDA_Graphs_Doc2025} could be used, by assigning each kernel call to a single independent node. This way the CPU would only make a single runtime call for the execution of all the kernels and the kernels should be able to overlap. However, adopting CUDA Graphs still increases programming complexity and poses run-time overhead. Even if the application is iterative, when the parameters of the kernels change dynamically on every iteration, there is some overhead to update them in the graph.
	
        In order to address the mentioned issues, we propose novel methodology that allows non-GPU programmers to automatically generate their own horizontally and vertically fused kernels for any algorithm, using a familiar high-level syntax. At the same time our methodology does not require GPU library providers to considerably increase development efforts writing the different possible combinations of vertically fused kernels, nor having to implement \glsxtrshort{hf} on every single kernel by hand. In order to accomplish this feat, our methodology defines an abstraction model and a common API, implemented using C++17 with our custom static polymorphism \cite{WG21_P2279R0} and static reflection. This API allows code from the same or different API-conformant GPU libraries to be easily fused into a fast single kernel, according to the final user requirements. Also, with this methodology we do not need to create a new compiler, since we leverage standard C++17 features, supported by most compilers such as \texttt{nvcc}.

	\section{Methodology}
\begin{figure}
		\centering		
			\includegraphics[keepaspectratio=true,width=0.48\textwidth]{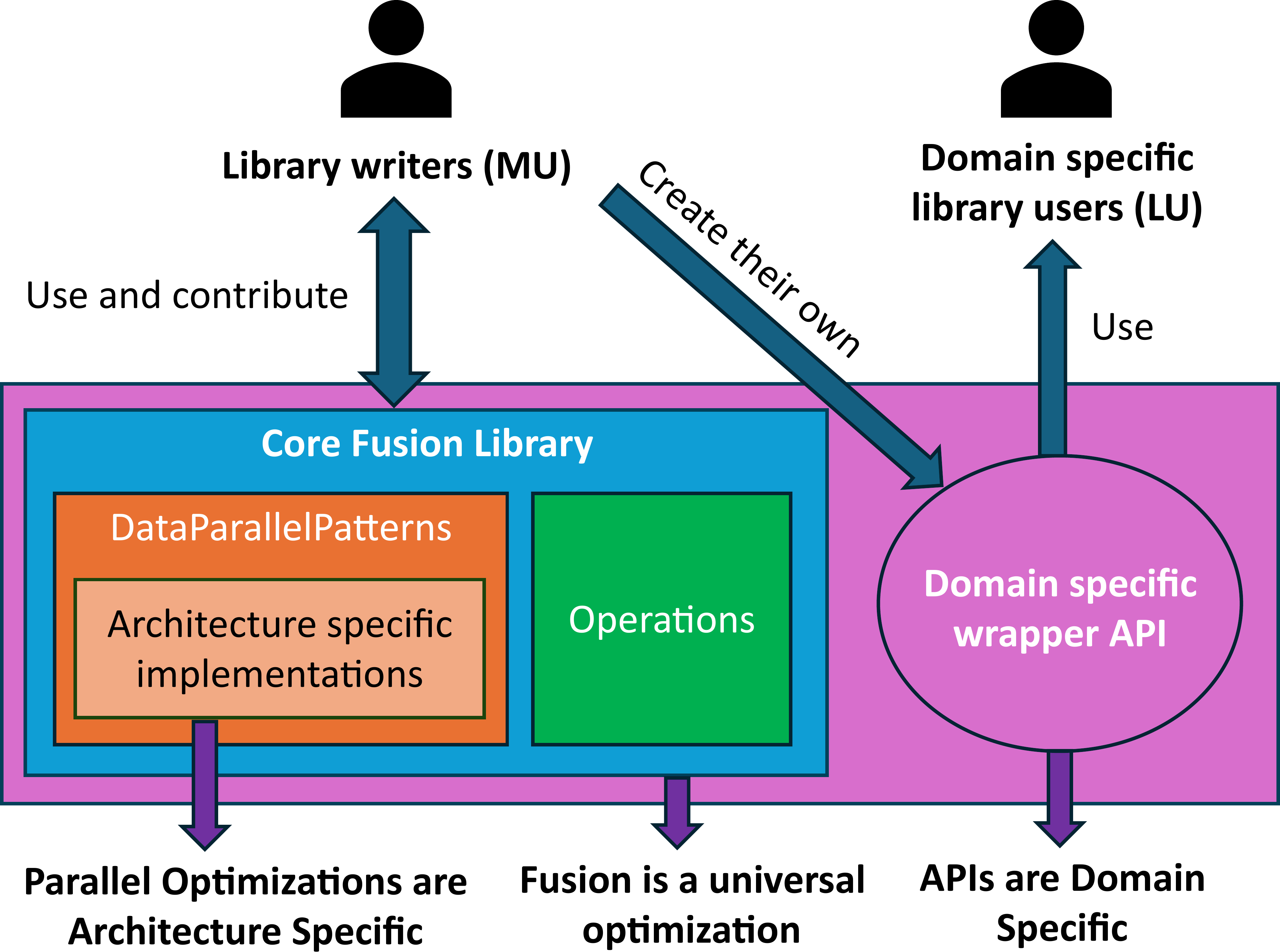}	
			\caption{ Representation of the interaction of the methodology and two types of final users. A single-core library may serve as the base for creating any derived domain-specific library, based on standard C++ syntax and compilers. }
			\label{fig:users}
		
	\end{figure}

        \noindent As depicted in Figure~\ref{fig:users}, we define domain-specific GPU library writers, as the \glsxtrfullpl{mu}, and non-GPU programmers in specific domains that use GPU libraries as the \glsxtrfullpl{lu}. We define a methodology for \glsxtrshortpl{mu} that enables them to create efficient GPU libraries. These libraries, following our methodology, enable the \glsxtrshortpl{lu} to perform automatic \glsxtrshort{vf} and \glsxtrshort{hf}, being completely oblivious to these terms.
    
\begin{figure}
		\centering
			\includegraphics[keepaspectratio=true,width=0.48\textwidth]{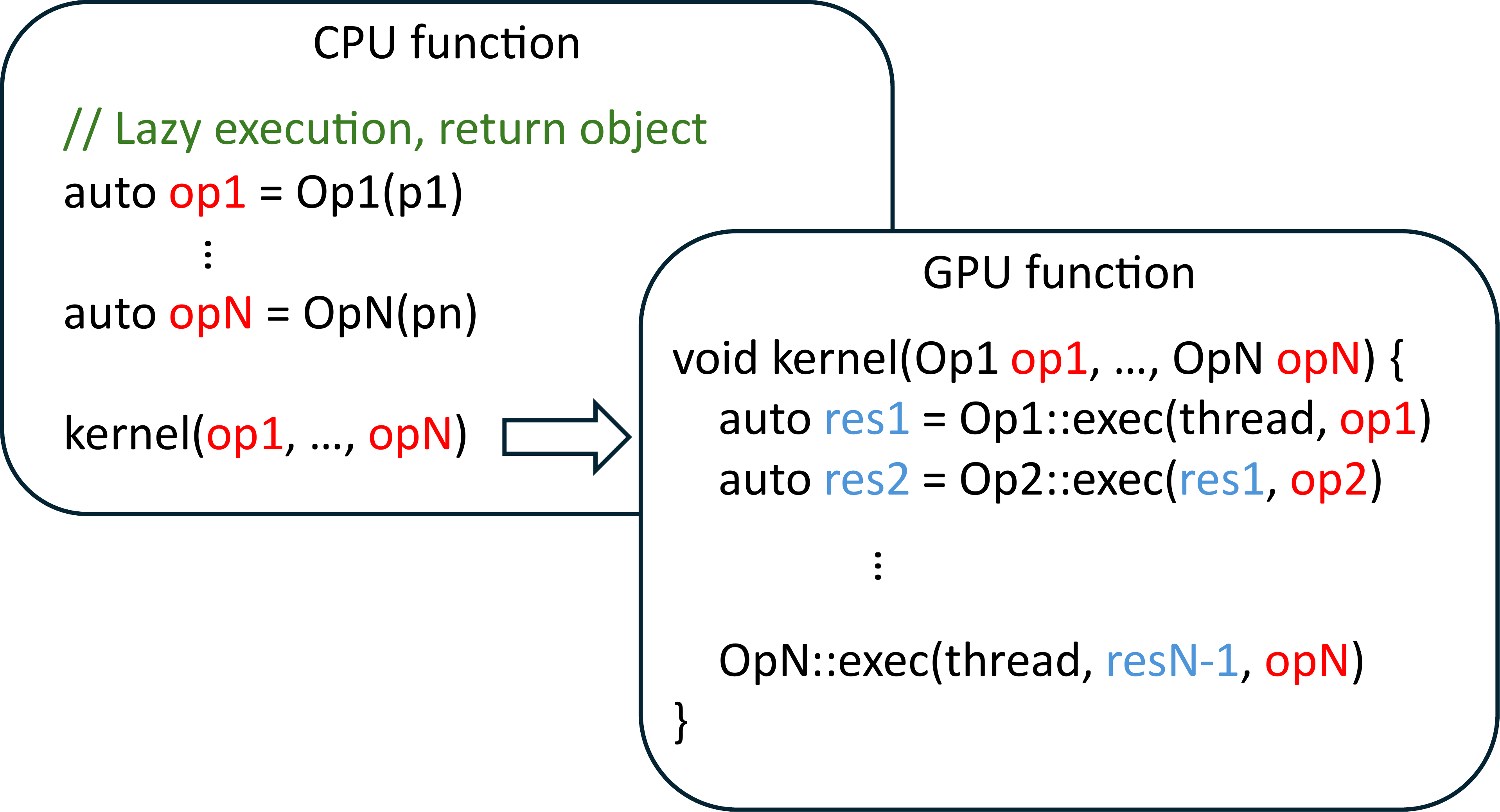}	
			\caption{ Figure showing the main idea behind the methodology. Defining a set of operations with the required parameters \texttt{pN} in CPU code, make them available inside a GPU kernel and execute them in the GPU kernel with a given set of dependencies and thread behavior. }
			\label{fig:kernelFromCPU}
		
	\end{figure}

        For the purpose of \glsxtrshort{vf}, we need to define GPU kernel components that are connectable inside a single kernel. In Figure~\ref{fig:kernelFromCPU} we illustrate the idea of making those components connectable (in red), referring to the possibility of connecting the output of one component to the input of the next component, as in a directed acyclic graph, or through some intermediate code. The data returned by one component, being the input for the following component, shall remain in registers (as in Figure~\ref{fig:kernelFromCPU}), preventing from the costly DRAM accesses that traditional libraries produce. That data shall be in registers under the following conditions, according to CUDA coding conventions:
	\begin{enumerate}
		\item The components have their executable code inside functions decorated with  \texttt{\_\_device\_\_}: Functions decorated with \texttt{\_\_device\_\_} may only be called inside the scope of a function decorated with \texttt{\_\_global\_\_} (a CUDA kernel) or another \texttt{\_\_device\_\_} decorated function.
        
		\item Both the input and output of those functions are non-pointer variables. This means that the input variables and the values returned will reside in GPU private memory (also known as registers or SRAM). Hence, each CUDA thread will feature its own private copy of the variables, with its own value.
	\end{enumerate}
    
        Therefore, our method revolves around connectable components that implement private memory operations. These connectable components cannot include thread behavior code, such as thread divergence or synchronization; any thread entering those components must execute the component code and produce a result. We call those components \glsxtrfullpl{op} and classify them in two main groups: \glsxtrfullpl{cop} and \glsxtrfullpl{mop}. 
	
        We define an additional component that will determine the thread behavior and will obtain a set of Operations as parameters, to execute these following the thread behavior it implements. We call these components \glsxtrfullpl{dpp}. In future work, we plan to be able to combine several \glsxtrshortpl{dpp} inside a kernel. In this paper we focus on a single Transform \glsxtrshort{dpp}, using several \glsxtrshortpl{op}.
	
	\subsection{Compute Operations}
	\label{sec:computeops}
	\glsxtrshortpl{cop} are the connectable components that implement the $K2$ step of a kernel as depicted in Figure~\ref{fig:verticalfusion}. We define two main types of \glsxtrshortpl{cop}:
	\begin{itemize}
		\item{\textbf{\glsxtrfull{uop}:} these are \glsxtrshortpl{cop} that only require the result of the previous \glsxtrshort{op} as input, in the shape of a single function parameter variable, as seen in Table \ref{tab:optypes}. It returns a single output variable as result. The types of the input and output variables are not restricted, including for example a tuple (\texttt{fk::Tuple}).}
		\item{ \textbf{\glsxtrfull{bop}:} these are \glsxtrshortpl{cop} that feature an additional function parameter (\texttt{params}), as exposed in Table~\ref{tab:optypes}. This parameter contains any values not generated by the kernel code and passed as parameters to the \texttt{\_\_global\_\_} function or kernel that is calling the \glsxtrshort{bop}.}
	\end{itemize}

\begin{table}
		\renewcommand{\arraystretch}{0.9}%
		\ttfamily 
		\fontsize{7.8pt}{10pt}\selectfont
		\bfseries
		\caption{{\tt exec} function versions according to the \texttt{OperationType}.}
		\label{tab:optypes}
		{
			\begin{tabularx}{\linewidth}{l l l}		
				\hline
				\textbf{OperationType }& K & exec() function signature \\
				\hline		
				\\
				\textcolor{teal}{ReadType} & K1 & \textcolor{teal}{OutputType} exec(\textcolor{blue}{const} fk::\textcolor{teal}{Point}\& thread, 
				\\ & & 
				\qquad \qquad  \qquad \qquad \quad\! \!\! \textcolor{blue}{const} \textcolor{teal}{ParamsType}\& params) \\[2ex]		
				\textcolor{teal}{BinaryType} & K2 & \textcolor{teal}{OutputType} exec(\textcolor{blue}{const} \textcolor{teal}{InputType}\& input,\\
				& & 
				\qquad 	\qquad\qquad\qquad\quad\! \!\!  \textcolor{blue}{const} \textcolor{teal}{ParamsType}\& params) \\[3ex]
				\textcolor{teal}{UnaryType} & K2 & \textcolor{teal}{OutputType} exec(\textcolor{blue}{const} \textcolor{teal}{InputType}\& input) \\[2ex]
				\textcolor{teal}{WriteType} & K3 & \,\,\,\,\qquad\quad \textcolor{blue}{void}  exec(\textcolor{blue}{const} fk::\textcolor{teal}{Point}\& thread,
				\\
				& & 
				\qquad \qquad \qquad\quad\quad\, \quad\textcolor{blue}{const} \textcolor{teal}{InputType}\& input,\\
				& & 
				\qquad \qquad \qquad\quad\quad\, \quad\textcolor{blue}{const} \textcolor{teal}{ParamsType}\& params) \\	
			\end{tabularx}
		}
	\end{table}
	
\begin{figure}
		\centering
		\begin{subfigure}{0.45\textwidth}
			\centering
			\includegraphics[keepaspectratio=true,width=1\textwidth]{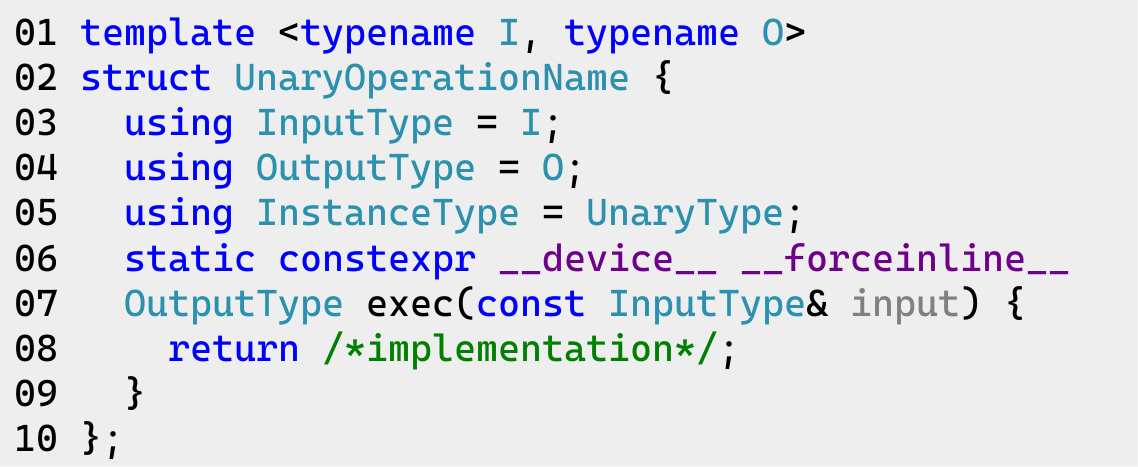}	
			\caption{ Unary Operation struct.}
			\label{fig:unaryoperation}
		\end{subfigure}\hfill
		
		\begin{subfigure}{0.45\textwidth}
			\centering	
			\includegraphics[keepaspectratio=true,width=1\textwidth]{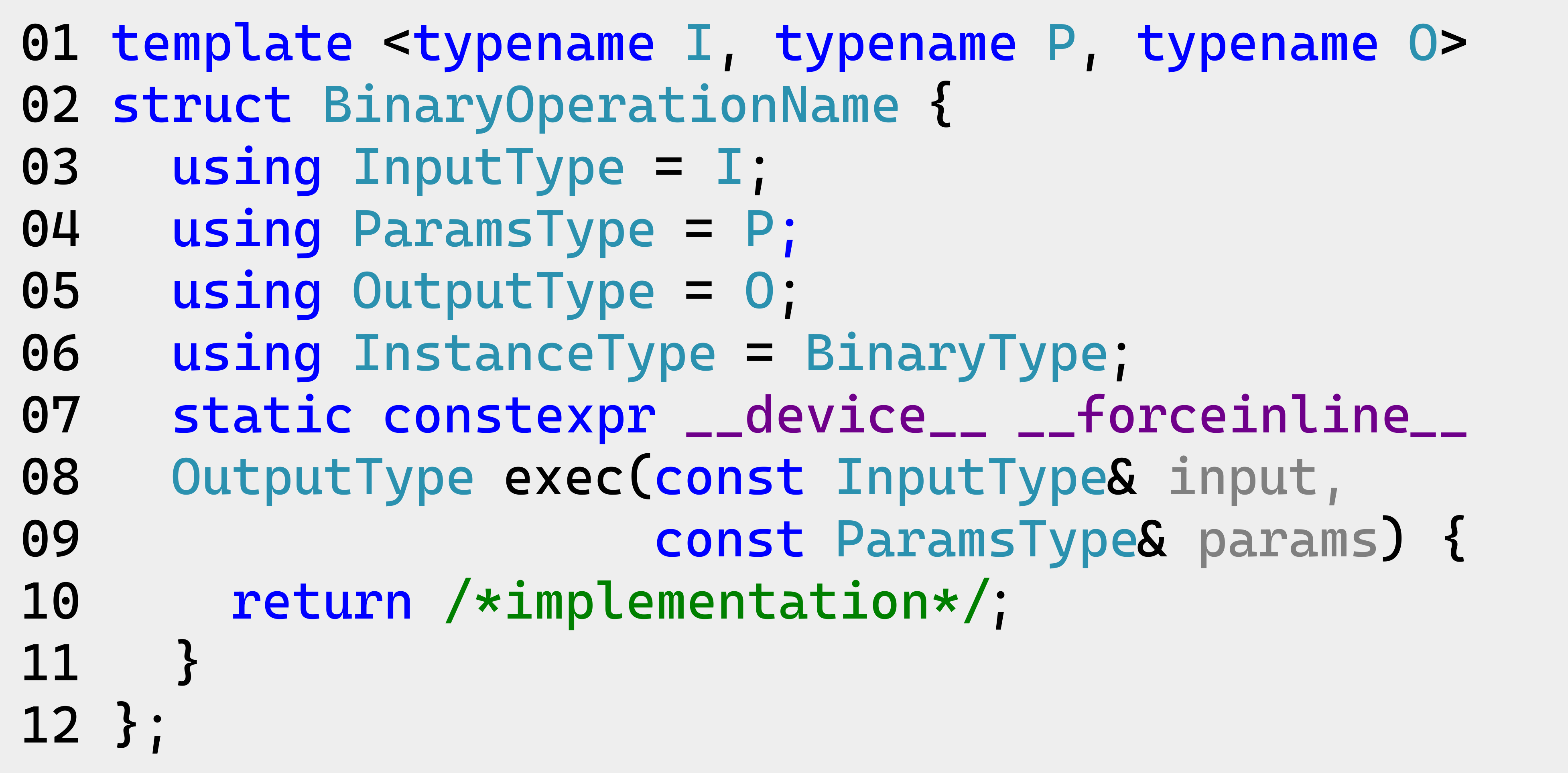}	
			\caption{ Binary Operation struct.}
			\label{fig:binaryoperation}
		\end{subfigure}
		\caption{Operation API in CUDA/C++ pseudo-code. The struct aliases are required to later be used with our custom static reflection techniques. In Figure~\ref{fig:unaryoperation} we see that all  \glsxtrshortpl{uop} must be implemented following this struct template: all \glsxtrshortpl{uop}  must be of type aliases \texttt{InputType},  \texttt{OutputType}, and  \texttt{InstanceType}. A \texttt{static \_\_device\_\_ \_\_forceinline\_\_} function named \texttt{exec} (\texttt{constexpr} is optional, depending on \texttt{exec} implementation and types) with return type \texttt{OutputType}, and first parameter type \texttt{const InputType\&} is also required. Likewise, the same concept applies to \glsxtrshortpl{bop} in Figure~\ref{fig:binaryoperation}, with the addition of type alias \texttt{ParamsType}, and \texttt{exec} function second parameter of type \texttt{const ParamsType\&}. } 
		\label{fig:UnaryBinaryOperation}
	\end{figure}

        In order for the user to be able to work with \glsxtrshortpl{uop} and \glsxtrshortpl{bop} and express an order of execution, we need runtime variables that represent those Ops. As shown in Figure~\ref{fig:UnaryBinaryOperation}, to represent the two Operation types, we use structs that do not contain any storage. This is done in purpose, so that Operations are strong types defining all the requirements of the Operation, but do not require to be instantiated at runtime in order to be used. This is especially interesting for \glsxtrshortpl{uop}  (Figure~\ref{fig:unaryoperation}).

\begin{figure}
		\centering
		\begin{subfigure}{0.4\textwidth}
			\centering
			\includegraphics[keepaspectratio=true,width=1\textwidth]{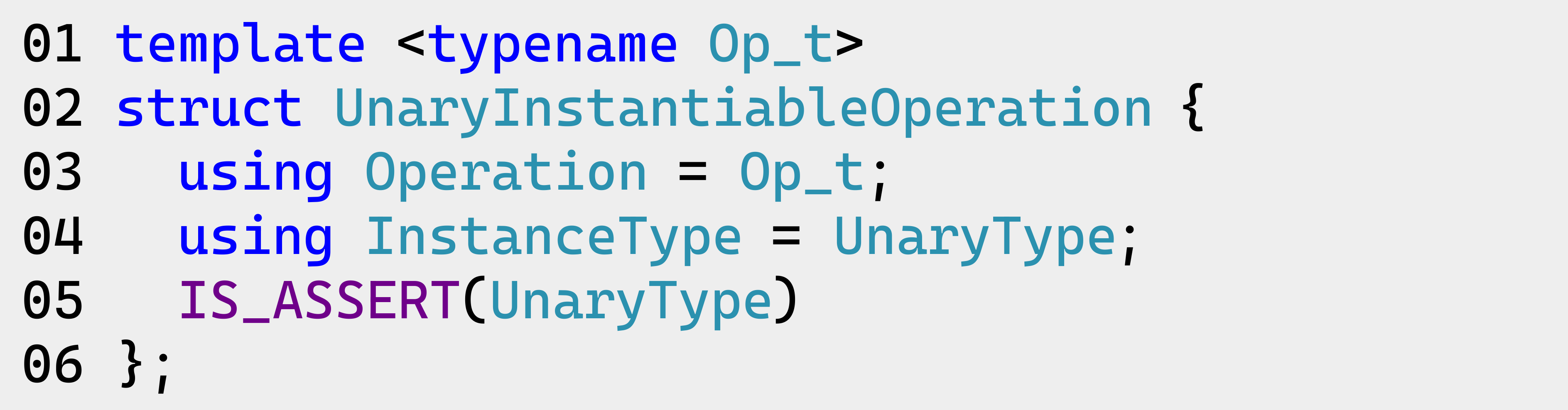}	
			\caption{ \glsxtrshort{uiop} struct. }
			\label{fig:UnaryInstantiableOperation}
		\end{subfigure}\hfill
		
		\begin{subfigure}{0.4\textwidth}
			\centering	
			\includegraphics[keepaspectratio=true,width=1\textwidth]{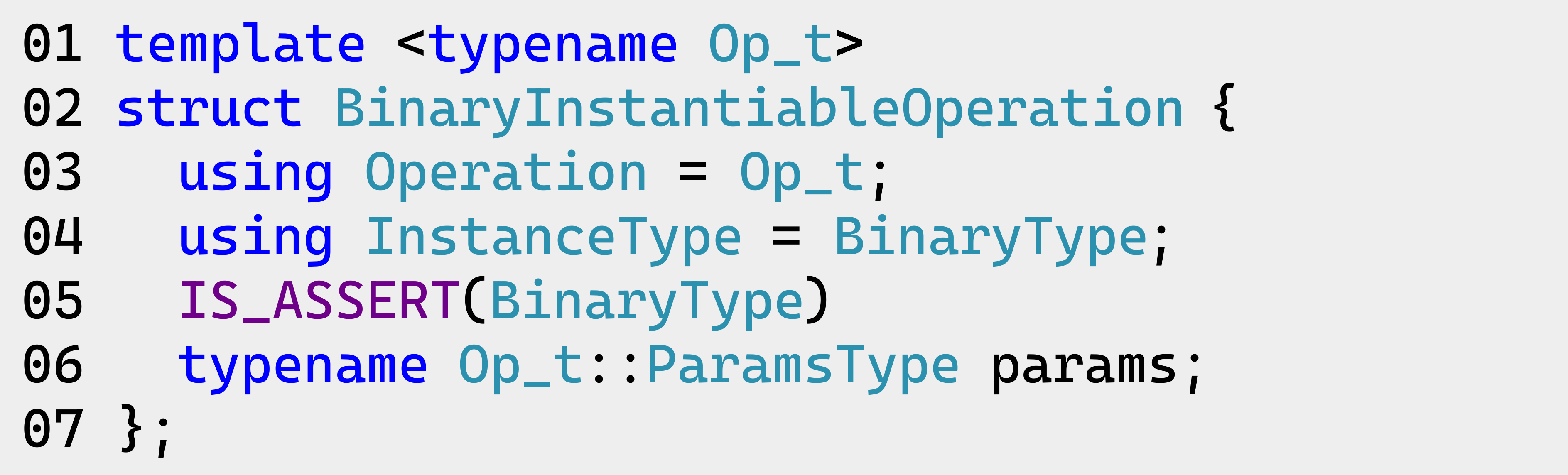}	
			\caption{ \glsxtrshort{biop} struct. }
			\label{fig:BinaryInstantiableOperation}
		\end{subfigure}
		
		\caption{ Pseudocode for \glsxtrfull{uiop} struct  and \glsxtrfull{biop}. } 
		\label{fig:InstantiableOperation}
		
	\end{figure}

        In order to have runtime variables with storage for \glsxtrshortpl{bop}, we define another set of structs intended only for that purpose, creating an instance of both \glsxtrshortpl{uop} and \glsxtrshortpl{bop} and storing the parameters in the case of  \glsxtrshortpl{bop} (Figure~\ref{fig:binaryoperation}). We call these structs \glsxtrfullpl{iop} and we show pseudo-code for their implementation in Figure~\ref{fig:InstantiableOperation}.
	
\begin{figure}
        \centering
        \includegraphics[keepaspectratio=true,width=0.48\textwidth]{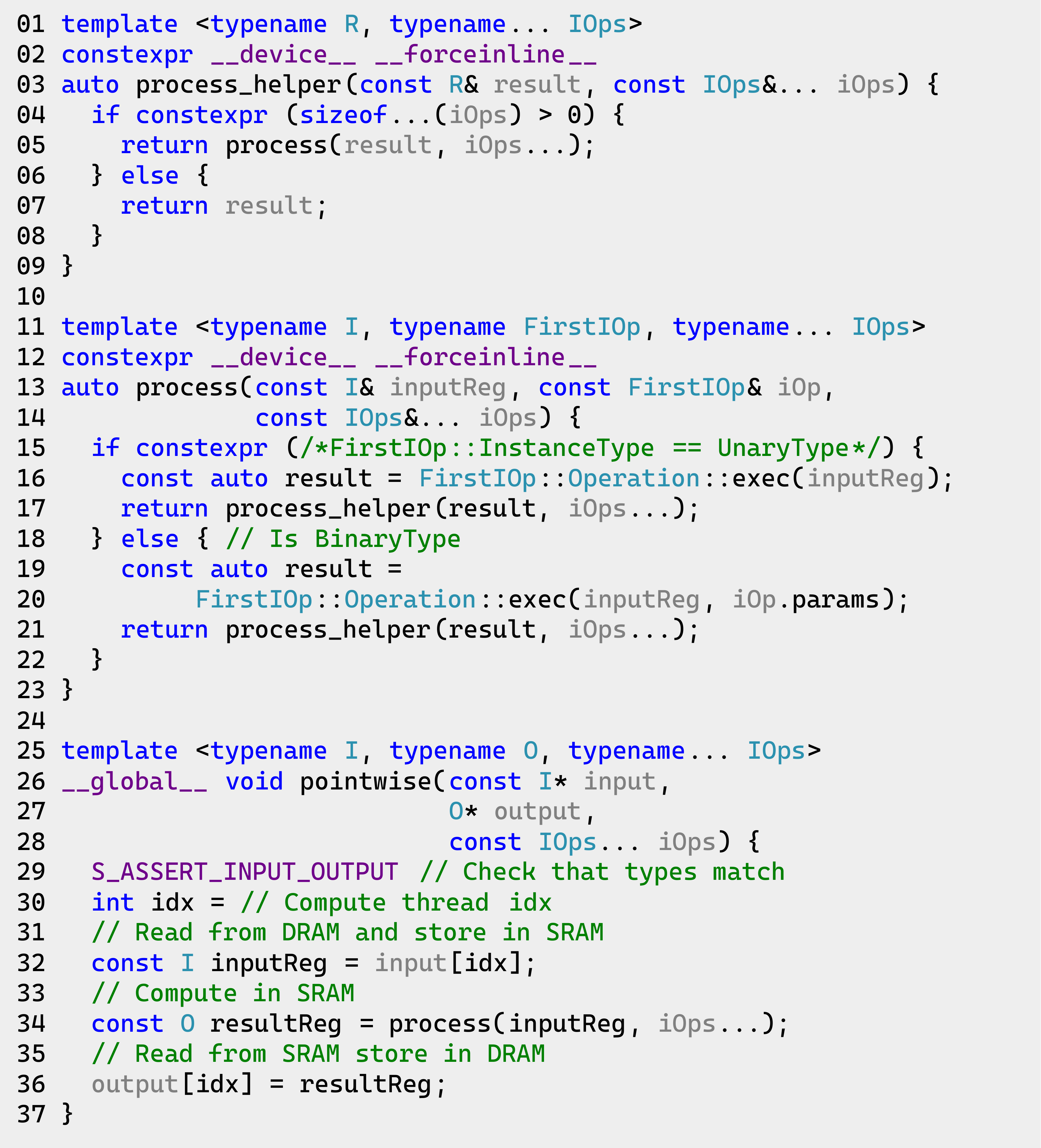}
        \caption{Example of implementation of a variadic CUDA kernel, capable of taking any number of Unary or Binary Operations. The read and write patterns are point-wise, in order to simplify this example. } 
        \label{fig:variadicKernel}
        \end{figure}

        With \glsxtrshortpl{iop} we may now create variables in host code, using the types represented in Figure~\ref{fig:InstantiableOperation}, which represent the GPU operations we are going to use. Next we may pass these variables as parameters of a variadic template \texttt{\_\_global\_\_} function, that will execute them in the same order they are passed, from left to right, as seen in Figure~\ref{fig:variadicKernel}. For simplicity, the code for reading and writing from/to GPU DRAM is hard-coded inside the \texttt{\_\_global\_\_} function.
	
	\subsection{Memory Operations} \label{sec:memops}
        \glsxtrshortpl{mop} are the connectable components that implement the $K1$ and $K2$ steps of a kernel as seen in Figure~\ref{fig:verticalfusion}. Following the GPU memory hierarchy specification, all GPU threads in a CUDA kernel have access to all the data in the pointers passed to the kernel, which reside in DRAM or device memory. This enables the CUDA threads to execute complex memory access patterns. Following the concept of connectable Operations we described in Section \ref{sec:computeops}, we again use {\tt structs} to define two main types of \glsxtrshortpl{mop}, as seen in Table \ref{tab:optypes}.

\begin{figure}
		\centering		\includegraphics[keepaspectratio=true,width=0.47\textwidth]{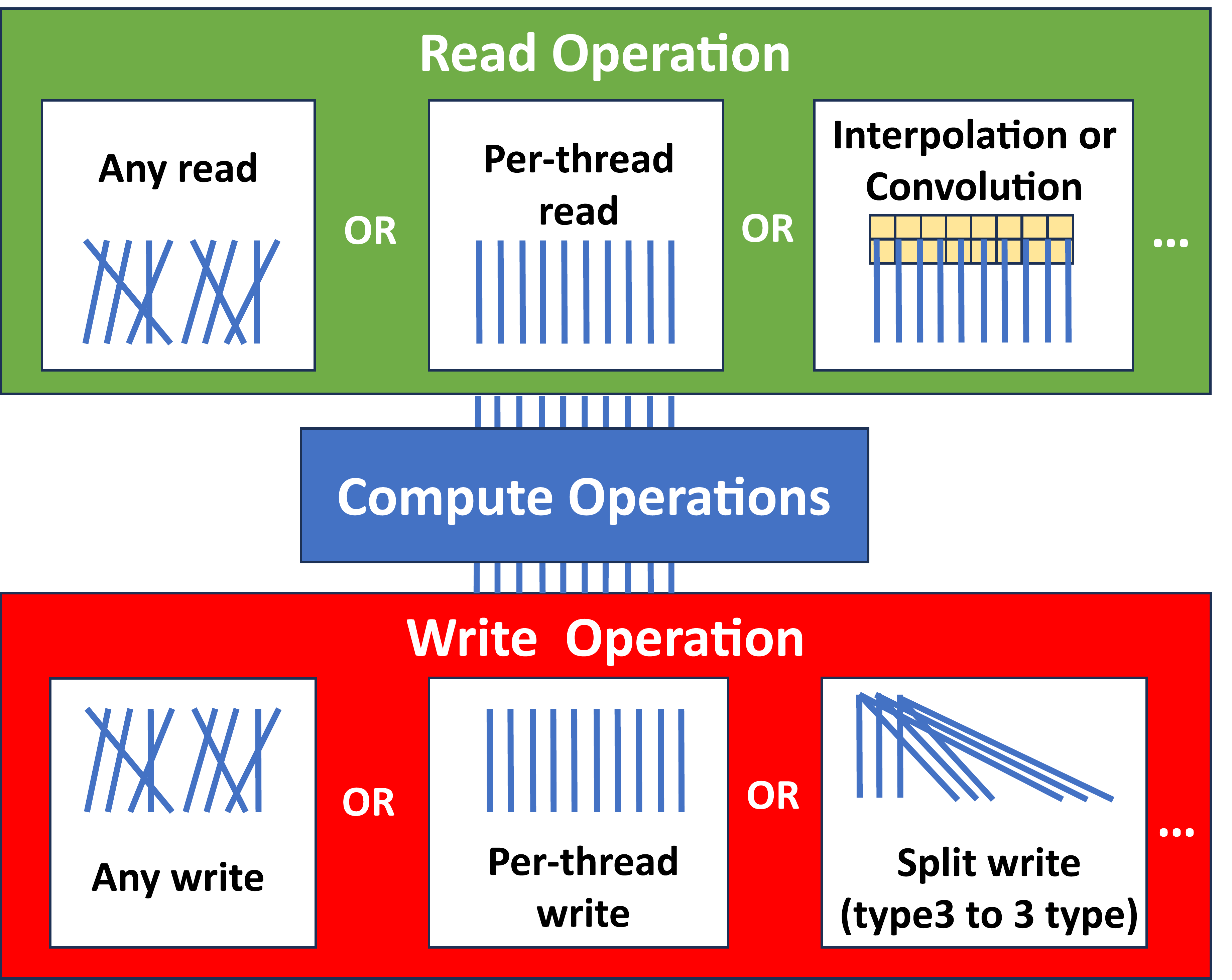}
		\caption{Graphical representation of several examples of \glsxtrshortpl{rop} and \glsxtrshortpl{wop}, depicting the CUDA threads with blue lines. When a blue line is not straight in a \glsxtrshort{mop}, it is reading or writing in a memory position index that does not directly map to the \glsxtrshort{tid}. The case of Split WOp, is an image processing operation that consists on transformig the packed pixel format of an image to planar pixel format. The reference in the image of type3 to 3 type represents any packed type with 3 channels, being stored as three image planes.} 
		\label{fig:MOps}
	\end{figure}

    Contrary to a \glsxtrshort{cop}, a \glsxtrshort{mop} may use \glsxtrfullpl{tid}  to compute the location to access in global memory, as shown in Figure~\ref{fig:MOps}. For instance, a thread may need to compute the location to read the data it needs using a combination of its \glsxtrshortpl{tid} and some values passed as parameters to the \glsxtrshort{mop}. According to the resulting value, the thread may even not read from DRAM and return a default value, or skip writing.

	The {\em Read Operation (ROp)} \glsxtrshort{mop} features an exec function without an input variable. Instead, it requires a \texttt{const fk::Point\& thread} variable that contains the {\tt (x,y,z)} coordinates that the Operation must use to compute the memory position or positions to read from DRAM. A second parameter (\texttt{const ParamsType\& params}) contains all the external parameters needed to perform the read operation, including the device pointer in DRAM. The {\tt exec} function returns the read data from DRAM in SRAM, to be used as input in the next Operation.
    
	The {\em Write Operation (WOp)} \glsxtrshort{mop}, on the other hand, features three parameters: the thread variable, the input variable with the data in SRAM, and a {\tt params} variable which includes the device pointer where the results will be written. To write the results in the correct memory position, the {\tt exec} function uses the thread variable coordinates and any additional information stored in {\tt params}, if required.
	
	Similarly to \glsxtrshortpl{cop}, we also define \glsxtrfullpl{riop} and \glsxtrfullpl{wiop}.

\begin{figure}
		\centering		
        \includegraphics[keepaspectratio=true,width=0.48\textwidth]{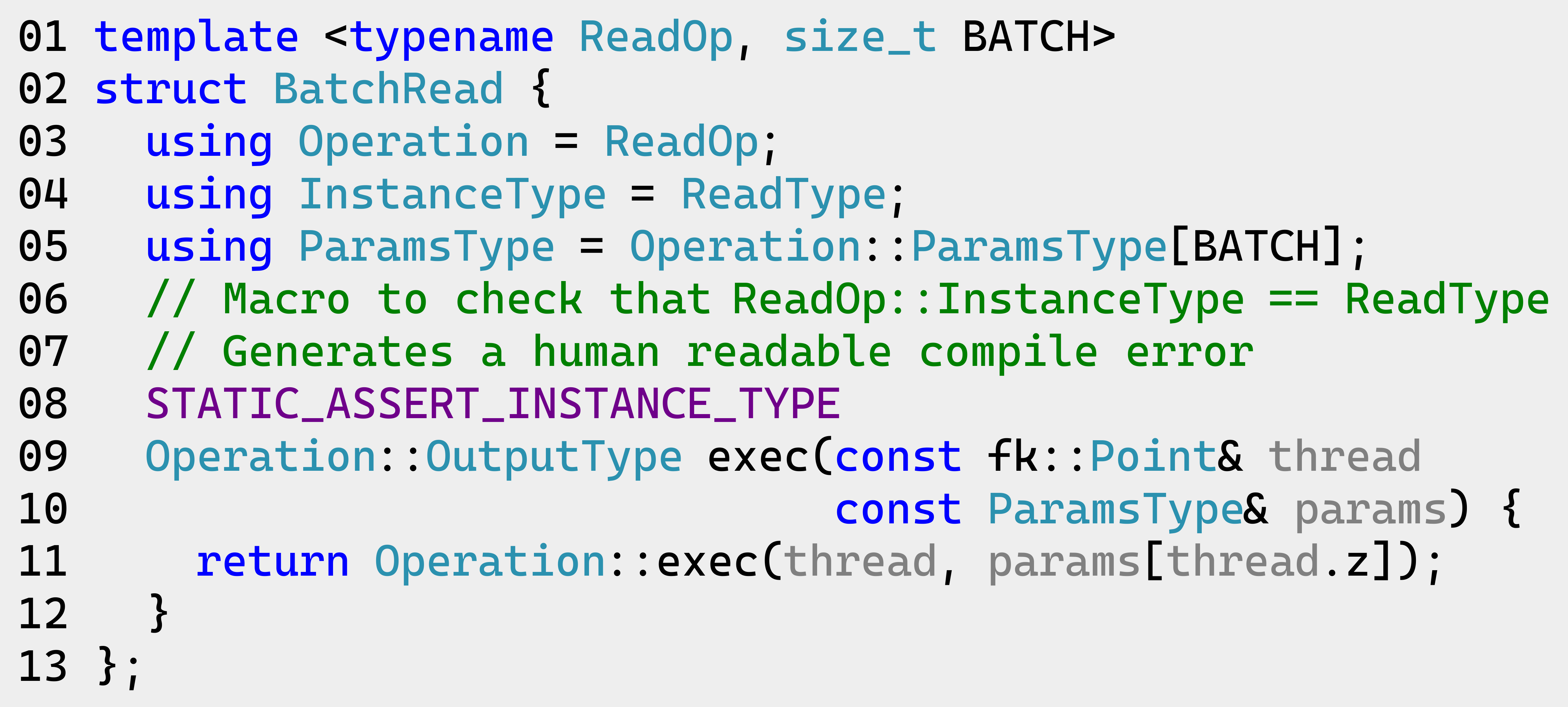}
		\caption{ Pseudocode of a \texttt{BatchRead} operation. } 
		\label{fig:BatchRead}
    \end{figure}

    In order to perform \glsxtrshort{hf}, we define two special types of \glsxtrshortpl{mop}, \texttt{BatchRead} (as seen in Figure~\ref{fig:BatchRead}) and \texttt{BatchWrite} (which is a \texttt{WriteType} \glsxtrshort{op} as seen in Table~\ref{tab:optypes}). These batch \glsxtrshortpl{op} require a \glsxtrshort{rop} or a \glsxtrshort{wop} and the number of times that \glsxtrshort{mop} needs to be called, both as template parameters, as seen in Figure~\ref{fig:BatchRead}. The batch memory operations feature as \texttt{ParamsType} an array of parameters and their \texttt{exec()} function is only responsible for calling \texttt{Operation::exec()} on the ParamsType array position determined by \texttt{blockIdx.z}.

 
    The \texttt{BatchRead} operation in Figure~\ref{fig:BatchRead} returns the result for the given \texttt{thread.z} and the following \glsxtrshortpl{cop} receives the corresponding input, according to \texttt{thread.z}. In the simplest case, the value of \texttt{thread.z} is \texttt{blockIdx.z}. Following this, any \glsxtrshort{mop} may use \glsxtrshort{hf}, by simply being the template parameter of the corresponding batch \glsxtrshort{op}.
    
    The dimensions of the grid, may be automatically inferred from data structures that contain the data dimensions. We implement our versions \texttt{RawPtr<ND, T>} and \texttt{Ptr<ND, T>\texttt}. If we obtain an \texttt{std::array<RawPtr<\_2D, float>, 30>} as input, we determine at compile time that we need to use \texttt{BatchRead<PerThreadRead<\_2D, float>, 30>}. If at every iteration the batch size is different, we still need to set the values in the non-used \texttt{thread.z} positions to a default value. We may accomplish this with a runtime variable, which will instruct the threads on the action to perform according to \texttt{thread.z} and that value.

	\subsection{Data Parallel Patterns}
	All the previous definitions are focused on per-thread \glsxtrshortpl{mop} and \glsxtrshortpl{cop}. These \glsxtrshortpl{op} may implement point-wise operations and any per-thread read and write pattern. But \glsxtrshortpl{cop} and \glsxtrshortpl{mop} do not allow to implement more complex thread behaviors such as reductions, optimized 2D and 3D convolutions, or optimized matrix multiplications, e.g.
	
	What these algorithms have in common, is that the number of threads being used is not constant throughout the execution, or does not directly map to the number of elements in the output data. In order to implement those algorithms we need to add thread divergence, iterations, and smart usage of shared memory and registers. To accomplish this goal we define an abstraction we call \glsxtrfull{dpp}. 
	
\begin{figure}
		\centering
		\includegraphics[keepaspectratio=true,width=0.48\textwidth]{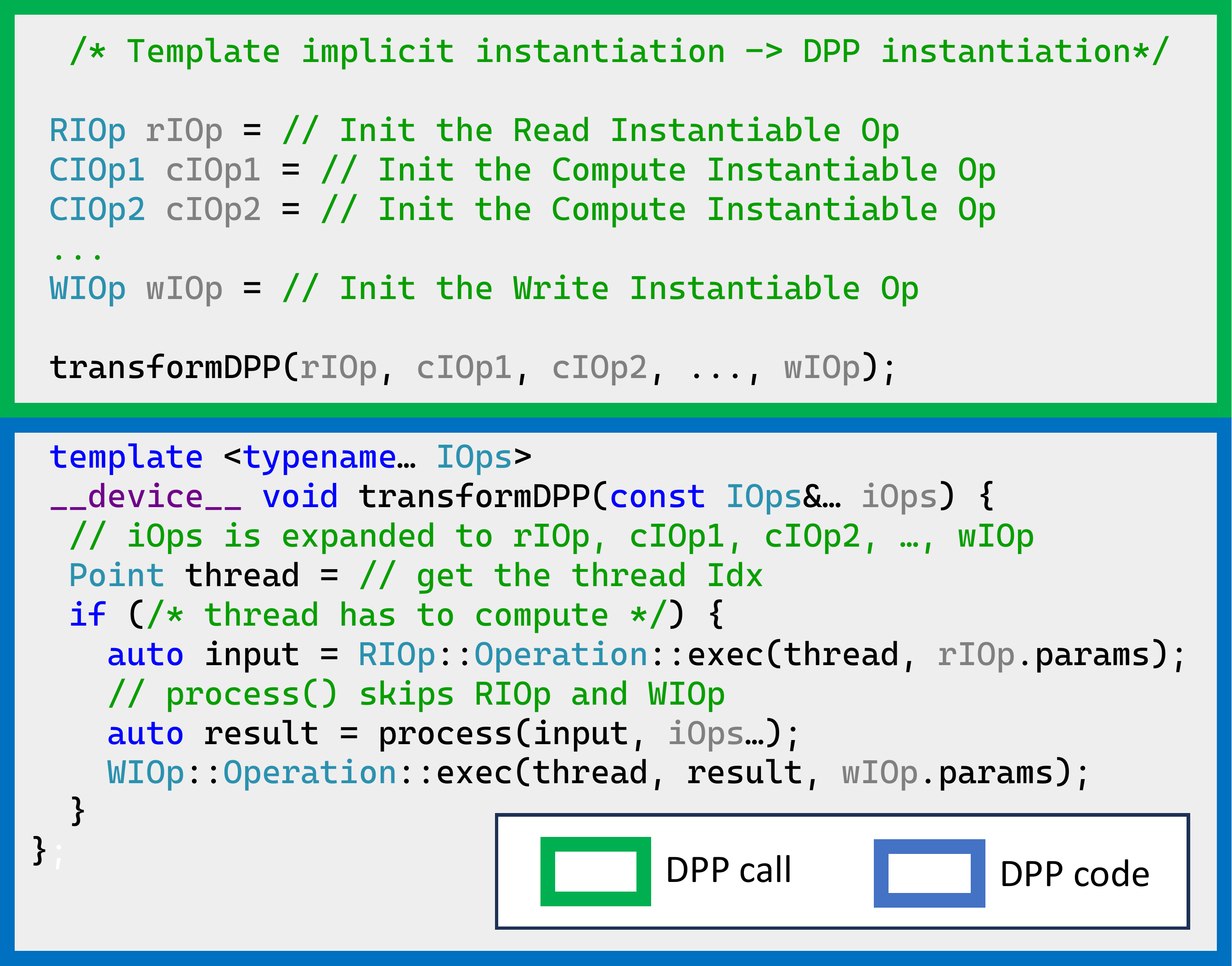}
		\caption{Example of implementation of a \texttt{TransformDPP}, which includes an optimization named \glsxtrfull{tc}. This consists on assigning further work to each thread by reading and writing with bigger data types. The code that belongs to the \glsxtrshort{dpp} is in blue and the code belonging to the Operations in orange. The number of \glsxtrshortpl{cop} passed to this \glsxtrshort{dpp} is variadic, which means that it is only limited by the template recursivity supported by the compiler. The \glsxtrshort{dpp} may deal with any number of \glsxtrshortpl{cop} by using recursive variadic template expansion, as seen in Figure~\ref{fig:variadicKernel}. } 
		\label{fig:TransformDPP}
	\end{figure}

	\glsxtrshortpl{dpp}, like \glsxtrshortpl{op}, are structs containing a \texttt{static \_\_device\_\_} function. In contrast to \glsxtrshortpl{op}, these functionn do not contain code that reads, modifies or writes input or output data. These only contain code that organizes the GPU threads and calls the {\tt exec} functions of the \glsxtrshortpl{op}, connecting them as the \glsxtrshort{dpp} requires. For instance, a \texttt{TransformDPP}, expects to receive as parameters a sequence of \glsxtrshortpl{iop}, as seen in Figure~\ref{fig:TransformDPP}, where the first \glsxtrshort{iop} is required to be a \glsxtrfull{riop}, and the last a \glsxtrfull{wiop}. Then, the \texttt{TransformDPP} will call the {\tt exec} function of the \glsxtrshort{riop}, and will pass the result as the input parameter of the {\tt exec} function of the next \glsxtrshort{iop}, the result of this \glsxtrshort{iop} to the next, and continue until the last \glsxtrshort{wiop}, which will write the results into the pointer contained in its parameters variable, as defined by its {\tt exec} function.

    In all this process, the only attributes that the \texttt{TransformDPP} needs about the \glsxtrshortpl{iop}, are the types and order of the \texttt{exec} function parameters, as seen in Table~\ref{tab:optypes}. Since all \glsxtrshortpl{iop} feature an alias called \texttt{InstanceType}, which determines whether the \glsxtrshort{iop} and corresponding \glsxtrshort{op} are {\tt Read}, {\tt Unary}, {\tt Binary}, or {\tt Write}, it knows which parameters to pass to each \texttt{exec} function. Using static reflection, the \glsxtrshort{dpp} may perform the \texttt{exec} selection at compile time, with an \texttt{if constexpr}, eliminating the need for runtime divergence.

\begin{figure}
		\centering
		\includegraphics[keepaspectratio=true,width=0.48\textwidth]{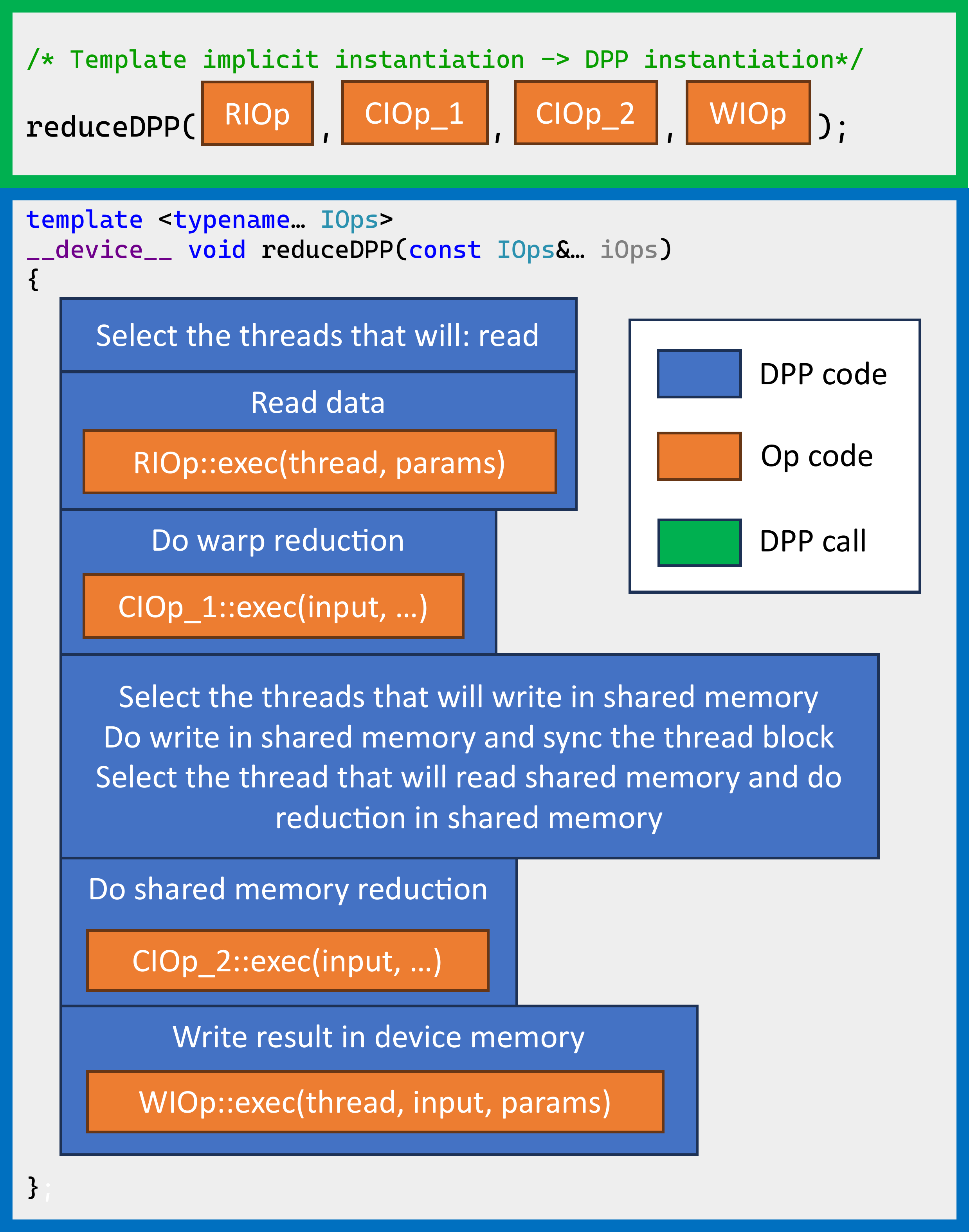}
		\caption{ Example of \glsxtrshort{dpp} (\texttt{ReduceDPP}) that implements a simplified version of a reduction with a single \glsxtrshort{tb}. The code that belongs to the \glsxtrshort{dpp} is in blue and the code belonging to the Operations in orange. The number of \glsxtrshortpl{cop} is limited to 2, since this is the number of operations required by this \glsxtrshort{dpp}. \texttt{CIOp\_1} and \texttt{CIOp\_2} may be the same or different \glsxtrshortpl{op}. } 
		\label{fig:ReduceDPP}
	\end{figure}

	In order to perform its duties, a \glsxtrshort{dpp} may need some information about the data dimensions and shape. This information is provided by the \texttt{ReadIOp}. With this information, the \glsxtrshort{dpp} performs the actions described in the inner blue boxes in figures~\ref{fig:TransformDPP} and~\ref{fig:ReduceDPP}.
	
	The \glsxtrshort{dpp} abstraction enables the implementation of several algorithms, depending on the \glsxtrshortpl{iop} passed as parameters. For instance, with a \texttt{ReduceDPP} that allows to pass more than one set of \glsxtrshortpl{iop}, for a given matrix we may find the maximum value, the minimum value, the addition of all the elements, and the mean value, all by reading the source data only once. Additionally, if new \glsxtrshortpl{gpu} incorporate new hardware features that accelerate a given \glsxtrshort{dpp}, the library programmers may update that \glsxtrshort{dpp}, and automatically all the usages of that \glsxtrshort{dpp} will benefit from the performance improvement. This eases code  maintainability, but it also eases its upgrade to new hardware.
	
	\subsection{The Final User Interface}
	\label{subsec:finalui}
	Using our methodology, we may provide the same functions as in other libraries, with almost the same parameters. The main difference is that the functions do not call any GPU kernel. Instead these return an \glsxtrshortpl{iop}, whose type and contents do not need to be known by the user. The execution of the \glsxtrshortpl{iop} will be performed at a later stage, as in lazy execution~\cite{functionalProgHughes}, when the user has defined all the chain of \glsxtrshortpl{iop}.
	
\begin{figure}
		\centering
		\includegraphics[keepaspectratio=true,width=0.38\textwidth]{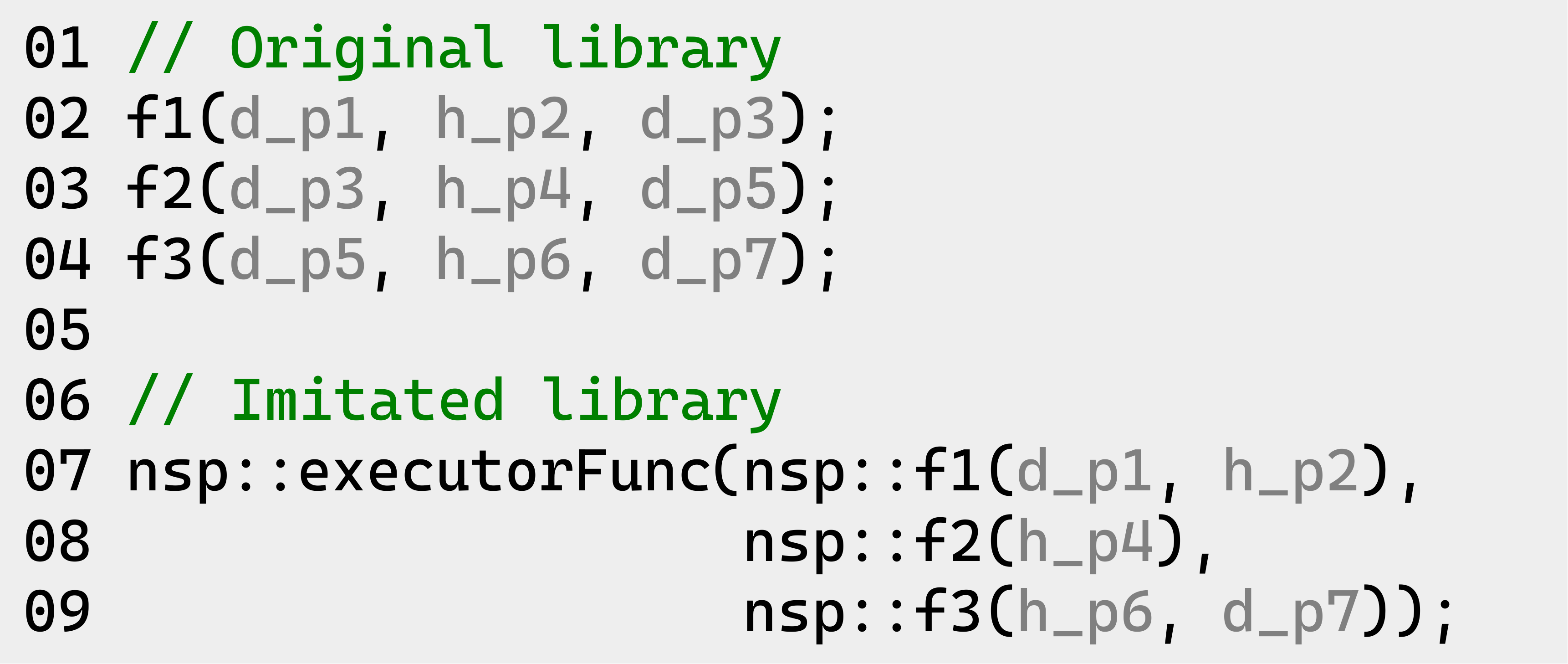}
		\caption{Abstract example of code syntax for an existing GPU library vs a wrapper of The Fused kernel Library that imitates that library. \texttt{nsp} is a namespace that will have been derived from the name of the wrapper. Variable names have a suffix \texttt{d\_} and \texttt{h\_} which indicate that the variable resides in device or host memory, respectively. Device variables will contain pointers to device memory, whereas host variables will contain non pointer values that will be copied to the GPU during the kernel call.} 
		\label{fig:FinalInterface}
	\end{figure}

	The user shall pass those \glsxtrshortpl{iop} as parameters of an executor function as in Figure~\ref{fig:FinalInterface} line 7. The \glsxtrshortpl{iop} types are used at compile time to generate the kernel code. The parameters stored inside the \glsxtrshortpl{iop} are used at runtime to execute the GPU kernel.
	
	Our replica of the original library functions is able to obtain the parameters using the same types as the original version. Internally, our functions move the data from the original library variables to our data types and instantiate the \glsxtrshort{iop} that will be returned.
	
	Since we are not storing intermediate results in DRAM, frequently our version of the functions will not require some of the DRAM pointers, effectively saving GPU memory and simplifying the code. In Figure~\ref{fig:FinalInterface} we can see how our version of functions \texttt{f1}, \texttt{f2}, and \texttt{f3}, does not need the variables \texttt{d\_p3} and \texttt{d\_p5}.

    With this highly similar syntax, we obtain speedups versus the original library from $\numprint{1}\times$ to $\numprint{16000}\times$, depending on the number of operations being fused and the size of the GPU. 
    The expected speedup is proportional to the number of kernels in the original code, starting at $\numprint{1}\times$ for a single-kernel call.
    
	\section{Experiments}
	\label{sec:experiments}
	\noindent In order to test our proposed methodology, we have implemented a \glsxtrshort{dpp} called \texttt{TransformDPP} and several \glsxtrshortpl{iop}. We have open-sourced our implementation as the \glsxtrfull{fkl}~\cite{fklrepo2025}. We have tested the FKL in several system configurations, as detailed in Table~\ref{tab:systems}. 
	
	On top of it, we have created two wrappers,  \glsxtrfull{cvgs} \cite{cvgsrepo2025}, which imitates the OpenCV-CUDA \cite{opencv_library,bradski2008learning,opencvrepo2025} library, and FastNPP \cite{fastnpprepo2025}, which imitates the NVIDIA Performance Primitives (NPP) library. The wrappers use a git submodule to the  \glsxtrshort{fkl}, since the actual functionality is in the \glsxtrshort{fkl} and the wrappers only translate OpenCV-CUDA or \glsxtrshort{npp} objects to \glsxtrshort{fkl} objects. With this, we demonstrate how simple the syntax can be, accomplishing one of our most important goals: enabling non-CUDA programmers to create their own fused and fast kernels, without CUDA knowledge and with a markedly softened learning curve.
	
	The goals of our experiments are two-fold:
	\begin{enumerate}
		\item To show that we obtain superior performance (lower execution times): with our strategy, compared to state--of--the--art CUDA based libraries.
		\item To show that we can attain this performance using a high-level API which is almost identical to that of the original library (as discussed in Section \ref{subsec:finalui}).
	\end{enumerate}
	
	We have implemented ten experiments where we compare the execution times of an external library $A$ with our implementation $B$. Where relevant we compare our implementation $B$ with $A$ using CUDA Graphs. On each experiment we explore the effects of changing only one of the possible parameters, to see the effects of that parameter on the resulting speedup obtained by our implementation $B$. For each parameter value, we repeat the experiment 100 times. On each time we record the execution time for A and B and compute the speedup ($A/B$). In the plots we either show the mean $A/B$ speedup or the mean execution times of $A$ and $B$. The maximum Relative Standard Deviation (RSD) for the series of 100 executions is highly dependent on the mean execution times. Experiments with less than 0.005ms of mean execution time, obtain RSD values from $8\%$ to a maximum of $25\%$. Executions with more than 0.005ms of mean execution times ($99\%$ of the cases) attain RSD values from $8\%$ to less than $0.01\%$. All the experiments, except for the "GPU size", are executed on an NVIDIA RTX 4090 GPU on Ubuntu 24.04 (system 5 in Table~\ref{tab:systems}).

	\begin{enumerate}
		\item {\textbf{\glsxtrshort{cvgs} wrapper overhead:} we compare identical kernels, executed using the \glsxtrshort{fkl}, with and without the \glsxtrshort{cvgs} wrapper. With this we assess if the \glsxtrshort{cvgs} wrapper poses any effect on performance, both in GPU and CPU.}
		
        \item {\textbf{Number of vertically fused operations:} we analyze the effects of increasing the number of vertically fused operations over the speedup. To that end, we use operations that lead to a single assembly instruction. The goal is to find the limit in which the speedup stops increasing.}
        
        \item {\textbf{Number of horizontally fused kernels:} we analyze how the number of horizontally fused kernels affects the speedup and compare it to the speedups obtained with \glsxtrshort{vf}.}
        
        \item {\textbf{Number of vertically fused operations including horizontal fusion:} in this test we combine \glsxtrshort{hf} and \glsxtrshort{vf} and assess how the speedup behaves when increasing the number of vertically fused operations.}
        
        \item {\textbf{Number of instructions per operation:} we fix a total amount of instructions to 500 and split them into 500 to 2 \glsxtrshortpl{op}. We turn each \glsxtrshort{op} into a single kernel and asses how the number of instructions per \glsxtrshortpl{op} affects the speedup when comparing to a single kernel with the 500 instructions.}
        
        \item {\textbf{CPU execution time:} we measure the CPU execution time when computing the kernel parameters before launching the kernels with our solution and compare it to OpenCV-CUDA and NPP.}
		
        \item {\textbf{Data size:} we measure how the input data size (number of elements), affects the speedup of our solution.}
        
		\item {\textbf{GPU size:} we use the experiment that produced the highest speedups and executed it on different GPU's with different FLOPS/B values. We asses if there is a clear correlation between FLOPS/B and the speedups obtained by using \glsxtrshort{vf} and \glsxtrshort{hf} combined.}
		
        \item {\textbf{Data type being used:} we measure how the speedups are affected by the data type.}
	
        \item {\textbf{Comparing to \glsxtrshort{npp}:} we compare the GPU performance of \glsxtrshort{npp} vs FastNPP.}
        \end{enumerate}
    
    \begin{table}
		\caption {Systems used to perform the tests. For desktop systems we execute the tests in Ubuntu Linux 24.04 with CUDA 12.8 and driver series 570. For embedded environments (Jetson AGX and Nano) we use Ubuntu 22.04, which comes with Jetpack 6.2. An additional server System (Grace-Hopper) has been tested only on Linux Ubuntu 24.04. The last row shows the number of floating point operations that the GPU can perform for every byte transferred.}
		{\renewcommand{\arraystretch}{1.2}
			
			\begin{tabularx}{\linewidth}{l l l l}
				\hline
				& \textbf{System 1} & \textbf{System 2} & \textbf{System 3}\\
				\hline
				\textbf{System Type} & Nano Super & Orin AGX & PC\\
				\textbf{CPU} & Cortex & Cortex & Ryzen 9\\
				          & A78AE & A78AE & 7945HX\\
				\textbf{ISA} & Armv8.2 & Armv8.2 & x86\_64\\
				\textbf{RAM (GB)} & 16 & 32 & 64\\			
				\textbf{GPU} & GA10B & GA10B & GA106\\ 
				\textbf{Compute cores} & \numprint{1024} & \numprint{2048} & \numprint{3328}\\
				\textbf{TFLOPS (FP32)} & 1.880 & 5.325 & 7.987\\
				\textbf{VRAM (GB)} & 16 (shared) & 32 (shared) & 12\\
				\textbf{Bandwidth (GB/s)} & 102.4 & 204.8 & 288\\
				\textbf{\glsxtrshort{flopb}} & 18.36 & 26 & 27.73\\					
			\end{tabularx}
		}
        {\renewcommand{\arraystretch}{1.2}
			\begin{tabularx}{\linewidth}{l l l l}
				\hline
				&  \textbf{System 4} &\textbf{System 5}\\
				\hline
				\textbf{System Type} & Grace-Hopper & PC \\
				\textbf{CPU}  & ARM Neoverse & Ryzen 7 \\
				          & V2 & 5800X3D  \\
				\textbf{ISA} & Armv9 & x86\_64  \\
				\textbf{RAM (GB)} & 480 & 128  \\			
				\textbf{GPU} & GH100 & AD102  \\ 
				\textbf{Compute cores} & \numprint{18432} & \numprint{16384} \\
				\textbf{TFLOPS (FP32)} & 62.08 & 82.58 \\
				\textbf{VRAM (GB)} & 96 & 24  \\
				\textbf{Bandwidth (GB/s)} & 900 & \numprint{1010}  \\
				\textbf{\glsxtrshort{flopb}} & 68.97 & 81.68  \\					
			\end{tabularx}
		}	
		\label{tab:systems}	
	\end{table}
    
    Additionally, we discuss the code syntax differences of OpenCV-CUDA versus \glsxtrshort{cvgs} and \glsxtrshort{npp} versus FastNNP, to prove that it is possible to obtain a highly similar syntax and high-level programming, with greatly increased GPU performance.

    Finally, we analyze another benefit from \glsxtrshort{vf}, which is GPU memory savings. Since we only read from the first input pointers and write only to the last output pointers, we do not need to allocate any intermediate GPU memory region.

	\section{Experimental evaluation}
	\label{sec:results}
    
	In this section we present and discuss the results of the experiments described in 
    Section \ref{sec:experiments}.
	
	\subsection{cvGPUSpeedup Wrapper Overhead}
        We have measured the CPU and GPU overheads added by the \glsxtrshort{cvgs} wrapper over the \glsxtrshort{fkl} and we found that it is negligible: the only additional task the wrapper performs is to copy the parameters from OpenCV-CUDA classes to \glsxtrshort{fkl} structs, without deep copying the pointers. In the case of the GPU, it is exactly the same code when using \glsxtrshort{cvgs} or \glsxtrshort{fkl}.
        
	\subsection{Number of Vertically Fused Operations}
        For this experiment, we have used matrices of size $\numprint{4096}\times \numprint{2160}$, where each element is a \texttt{uchar}. The number of matrices or batch size used is 1; therefore, in this experiment we are not using \glsxtrshort{hf}, only \glsxtrshort{vf}. Given the size of the matrix, \glsxtrshort{hf} would not be beneficial. The kernel configuration is $32 \times 8 \times 1$ threads per thread block, and $128\times 270\times 1$ thread blocks in the grid.
        We have executed the experiment using two different sets of operations:
        \begin{itemize}
		\item {\textbf{Mul and Mul Operations}: the first benchmark creates a Vertically Fused kernel, using \glsxtrshort{cvgs}, with 2 consecutive multiplication operations. We compute the speedup of a kernel with the two operations inside, compared to launching two consecutive kernels that perform only a single Mul. Then, we repeat the experiment increasing the number of consecutive operations by 100 and compute the speedup again, this time comparing a single kernel with 102 operations inside, with launching 102 kernels with a single Mul inside. We continue increasing the number of Mul operations by 100, until \numprint{19902} operations.}
        \item {\textbf{Mul and Add Operations}: the second benchmark creates a first vertically fused kernel with 2 consecutive operations that consist in multiplication and addition. Then we increase the total number of operations by 100, by adding 50 pairs of Mul and Add operations. As in the first experiment, we continue increasing the number of fused operations until we attain the maximum speedup.}
        \end{itemize}

        In both cases, we are using a single instance of each \glsxtrshort{op} type, and an special \glsxtrshort{op} we called \texttt{StaticLoop}, in order to avoid consuming the parameter space of the kernel. The goal is to assess the best scenario, where the assembly is made of one operation after the other, without any other operation in between. We repeated the two experiments using CUDA Graphs with OpenCV-CUDA, to evaluate how CUDA Graphs compares with our implementation of \glsxtrshort{vf}.
        
        With this experiment, we can see that, given that the operations are inline functions carefully implemented to maximize the chances of in-lining, the compiler obtains all the code in a single scope. In that context, the compiler can fuse multiply and add instructions and perform other optimizations. We have verified this is indeed the case by inspecting the resulting SASS code using NSight Compute.
        
\begin{figure}
		\centering
		\includegraphics[keepaspectratio=true,width=0.48\textwidth]{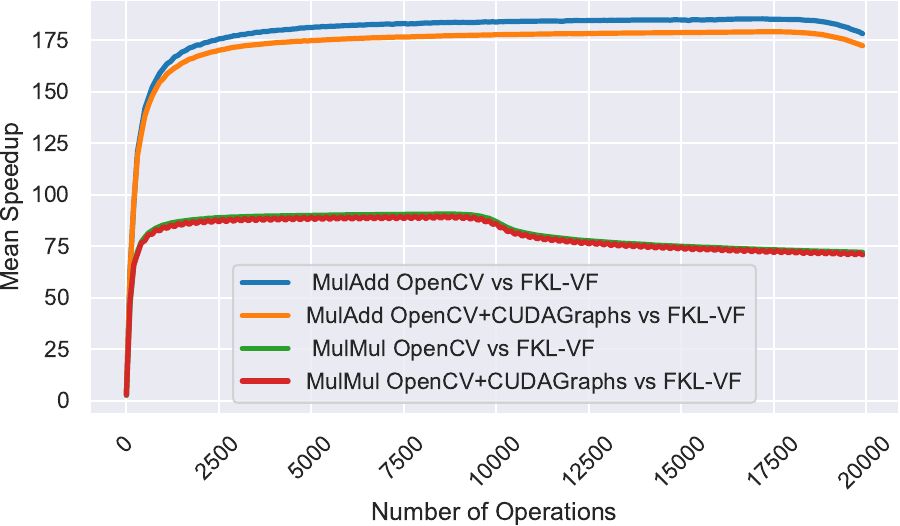}        
		\caption{VF-only experiment: speedup of \glsxtrshort{cvgs} versus OpenCV-CUDA, executing the same chain of $N$ operations, with $batch=1$ and matrix resolution $\numprint{4096}\times \numprint{2160}$, where $N$ is from 2 to \numprint{19902}.} 
		\label{fig:VerticalOnly}
	\end{figure}

        In Figure~\ref{fig:VerticalOnly} we can see that for the same total number of operations being executed, the combined {\tt Mul+Add} kernel is twice as fast as its counterpart. This is because, as mentioned, the compiler can fuse instructions and use the {\tt FMADD} instruction to perform the multiplication and addition. The maximum speedup of cvGS versus the OpenCV-CUDA baseline with pairs of Mul and Add ($185\times$) is approximately twice as much as that with pairs of Mul and Mul ($90\times$).

        As we can see in Figure~\ref{fig:VerticalOnly}, using CUDA Graphs provides a marginal improvement over OpenCV-CUDA with streams. This is consistent with the fact that CUDA Graphs does not perform \glsxtrshort{vf}; it still launches individual kernels one after another, leading to the same number of DRAM reads and writes as with OpenCV-CUDA alone. The small speedup decrease observed in the figure when using CUDA Graphs is due to the fact that with this solution the CPU only calls the CUDA runtime once, and the rest is handled directly by the GPU, hence reducing the time spent among kernel calls.
        
        \subsection{Number of Horizontally Fused Kernels}
        For this experiment we benchmark for batches of 10 to 600 images of $60\times 120$ \texttt{uchar} elements each. 
        The \glsxtrshortpl{op}. to perform are: $Batch(Read\rightarrow Cast \rightarrow Multiply \rightarrow   Subtract\rightarrow Divide\rightarrow Write)$.
        To avoid mixing the speedups with \glsxtrshort{vf}, we use \glsxtrshort{cvgs} with \glsxtrshort{vf} in both cases. One of the executions additionally performs \glsxtrshort{hf} and the other executes \glsxtrshort{vf} kernels only in a loop.

        In order to avoid measuring the influence of the CPU part that handles arrays of parameters, we generate the \glsxtrshortpl{iop} before starting to measure the execution time of the kernels. We repeat the experiment by launching the individual \glsxtrshort{vf} kernels generated with \glsxtrshort{cvgs} using CUDA Graphs, to evaluate the impact it poses on \glsxtrshort{hf}.

\begin{figure}
		\centering        \includegraphics[keepaspectratio=true,width=0.49\textwidth]{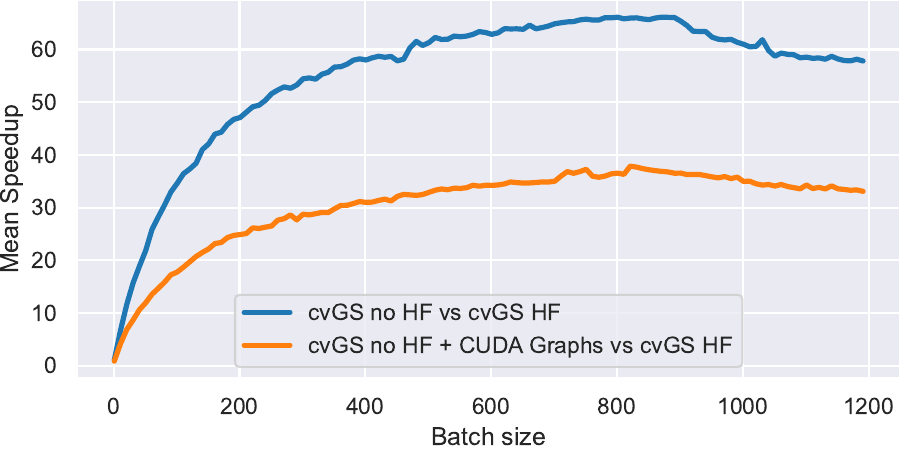}
		\caption{HF-only experiment: speedup of calling a kernel for each batch element versus calling a single kernel using \glsxtrshort{hf}, with a matrix size of $60\times 120$ with \texttt{uchar} data elements, and batch sizes from 1 to \numprint{1191}.} 
		\label{fig:ResultsHorizontalFusion}
	\end{figure}

        The kernel configuration is $32\times 8\times 1$ threads per thread-block and $2\times 15\times N$ thread blocks in the grid, where $N$ is from 1 to \numprint{1191}.
        As we can see in Figure~\ref{fig:ResultsHorizontalFusion}, the speedup grows steep for small batch numbers and it slowly decelerates for increasing batch sizes, since the benefit from using more resources decreases when we approach the GPU resource limits with larger batch sizes. We have confirmed with NSight Compute that the kernel used in this experiment is \glsxtrshort{mb} (i.e., the limiting resource in this case is the memory bandwidth).

        We observe a maximum speedup of $66\times$ when comparing \glsxtrshort{cvgs} without using \glsxtrshort{hf} to \glsxtrshort{cvgs} with \glsxtrshort{hf} and a maximum speedup of $37\times$ when comparing \glsxtrshort{cvgs} using \glsxtrshort{hf} via CUDA Graphs versus \glsxtrshort{cvgs} with our own \glsxtrshort{hf}. These results, along with the \glsxtrshort{vf} experiment, reveal that the speedup that comes from \glsxtrshort{vf} may be greater than that coming from \glsxtrshort{hf}, and optimization constructs such as CUDA Graphs only help if there is a chance for \glsxtrshort{hf}.

        \subsection{Number of Vertically Fused Operations Including HF}
        \label{sec:vfplushf}
        In this experiment, we leverage the fastest operation pair from the \glsxtrshort{vf} experiment, {\tt Mul+Add} and $batch=50$. We maintain the \texttt{uchar} data type for the matrix elements of $60\times 120$. 

        The \glsxtrshort{cvgs} version performs automatic \glsxtrshort{vf} and \glsxtrshort{hf}, resulting in a highly efficient single kernel, while the OpenCV-CUDA version launches one kernel for every operation and every batch element. The kernel configuration is $32\times 8\times 1$ threads per thread block, with $2\times 15\times 50$ thread blocks in the grid.
        We repeat this experiment using OpenCV-CUDA with CUDA Graphs vs cvGS, by creating a \texttt{cudaGraphNode\_t} per iteration and adding each node as a source node of a global \texttt{cudaGraph\_t}.
  
\begin{figure}
		\centering		\includegraphics[keepaspectratio=true,width=0.48\textwidth]{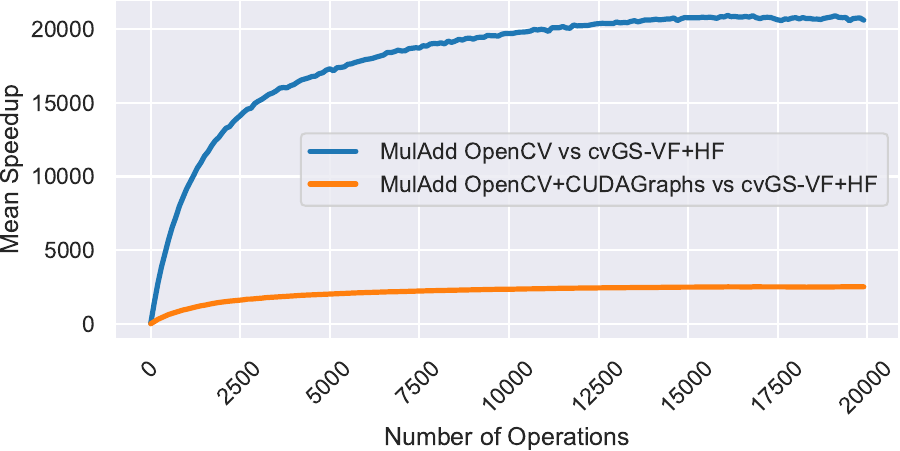}
		\caption{Combined VF and HF experiment: speedup of \glsxtrshort{cvgs} versus OpenCV-CUDA (including CUDA Graphs). Executing 1--\numprint{10000} pairs of {\tt Mul+Add} operations with a batch size of 50.} 
		\label{fig:VerticalHorizontalFusion}
	\end{figure}

        In Figure~\ref{fig:VerticalHorizontalFusion} we can see that the speedup resembles a logarithmic function. The maximum speedup is $\numprint{20931}\times$ faster with our method than with OpenCV-CUDA.
    
        The initial acceleration in speedup growth is explained by the fact that the kernel is \glsxtrshort{mb}. Therefore, adding more consecutive operations than one, does not increase the execution times of the kernel. When the kernel becomes a \glsxtrshort{cb}, adding more operations does increase the execution time but only by half a cycle per operation per warp. This is because the pair of operations {\tt Mul} and {\tt Add} become a single instruction. Instead, in the case of the individual kernels, adding a new kernel brings along all the time needed to read and write the data, which is substantially higher.
    
        The speedup stops growing because the generated code does not use any loop. Since it is generated with template recursion, the amount of code keeps growing. When the assembler code of a kernel reaches a certain number of lines, the GPU scheduler may decide to change the number of \glsxtrshortpl{tb} per streaming multiprocessor to schedule, or even to perform register spilling, in order to store part of the code in device memory. This increases the execution time of the kernel.

        In the case of CUDA Graphs, we can see in Figure~\ref{fig:VerticalHorizontalFusion} that in this case the speedup of our solution is substantially greater. This occurs because CUDA Graphs does not perform \glsxtrshort{vf}, where we obtained a $175\times$ speedup at 5000 \glsxtrshortpl{op}. Additionally, the \glsxtrshort{hf} improvement provided by CUDA Graphs is inferior to our \glsxtrshort{hf} solution, where we attained a $12\times $ speedup over CUDA Graphs at batch size 50. As a consequence, the maximum combined \glsxtrshort{vf} $\times$ \glsxtrshort{hf} speedup we obtained with respect to OpenCV-CUDA + CUDA Graphs is $\numprint{2527}\times$, close to a direct speedup multiplication of $12 \times 175 = 2100$. The additional speedup we obtained, can be explained by the fact that CUDA Graphs has to handle the launch of $50 \times 5000$ CUDA kernels, versus a single kernel with our implementation, but unfortunately, it is not possible to profile the CUDA Graphs runtime to validate our hypothesis.

        It is relevant to note that the CUDA Graphs implementations require an understanding of the concepts of \glsxtrshort{vf} and \glsxtrshort{hf} in order to properly develop them, heavily increasing the implementation complexity compared to our imitation of OpenCV-CUDA, cvGS. With cvGS, we are providing automatic \glsxtrshort{vf} and \glsxtrshort{hf} without requiring any special learning from the final user and much greater performance than CUDA Graphs. We explore the code syntax aspect in Section~\ref{exp_api}.
    
	\subsection{Number of Instructions per Operation}
        In this experiment we evaluate the speedup changes when \glsxtrshort{vf} is performed with \glsxtrshortpl{op} that leverage an increasing number of instructions. We compare the execution time of a single kernel with all the instructions with that of executing a kernel for each \glsxtrshort{op}. On each execution the number of \glsxtrshortpl{op} is gradually reduced, since each \glsxtrshort{op} execute further instructions. We set the total number of instructions to be executed in both cases constant to 500.

        The number of instructions per operation starts with 1 {\tt Mul} \glsxtrshort{op} with {\tt float} input and oputput data types. The number of operations is incremented by 5 on each execution, until 496. To reach 500 instructions executed, the last \glsxtrshort{op} contains the remaining instructions, when 500 is not divisible by the number of instructions per operation.
        The kernel configuration is $256\times 1\times 1$ threads per thread block and $\numprint{259200}\times 1\times 1$ thread blocks in the grid.
        
\begin{figure}
		\centering
        \includegraphics[keepaspectratio=true,width=0.485\textwidth]{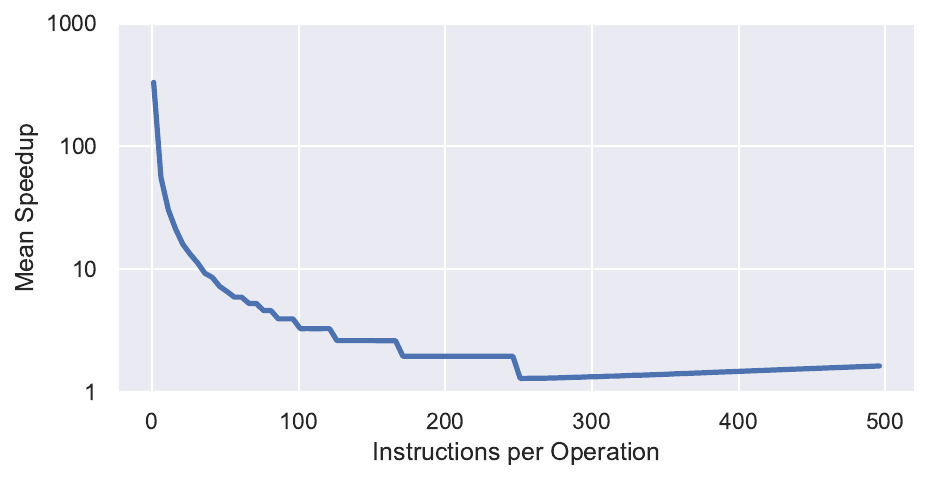}        
		\caption{Number of instructions per operation: logarithmic scale speedup of executing 500 instructions fused in a single kernel versus launching a varying number $N$ of kernels with a single \glsxtrshort{op} that contains a varying number $M$ of instructions, where always $N + M = 500$.} 
		\label{fig:NumInstructionsPerOperation}
	\end{figure}

        As we can see in Figure~\ref{fig:NumInstructionsPerOperation}, the speedup decreases when the number of instructions per \glsxtrshort{op} increases. Since having more instructions per operation entails executing less kernels, executing less kernels for the same number of instructions produces less reads and writes to DRAM comparatively with our approach, which leads to shorter execution time, as we have been discussing across the paper.

        In Figure~\ref{fig:NumInstructionsPerOperation} we also see some points where the speedup suddenly diminishes. These points reflect where the number of kernels (and therefore \glsxtrshortpl{iop}) is reduced by one. After that point, we see a stable speedup, until 250 instructions per Operation, where the speedup starts to increase. That speedup increase is due to the fact that the last kernel executes a number of instructions sufficiently small to turn it into \glsxtrshort{mb}, while the other kernels are \glsxtrshort{cb}. This means that when we increase the number of instructions in the \glsxtrshort{cb} kernels, these increase their execution time, while the \glsxtrshort{mb} kernel does not reduce its execution time, despite executing less instructions. This is another confirmation of the effects of the GPU latency hiding.
        
	\subsection{CPU Execution Time}
        In this experiment, we assess the impact that our methodology poses on CPU execution time. Considering the reduction in the number of times we call the CUDA runtime (in our case, CUDA kernel calls), we expect considerable improvement in total CPU execution time.
        The operations performed in this experiment are: $Batch(Crop \rightarrow Resize \rightarrow ColorConvert \rightarrow Multiply \rightarrow Subtract \rightarrow Divide \rightarrow Split)$. Resized images are $64\times 128$ pixels.

        For this fixed chain of Operations, we vary the batch size from 2 to 152 and observe how the speedup is affected. The kernel configuration is $32\times 8\times 1$ threads per thread block and $2\times 16 \times N$ thread blocks in the grid, where $N$ is from 2 to 152 thread blocks.

\begin{figure}
		\centering
        \includegraphics[keepaspectratio=true,width=0.49\textwidth]{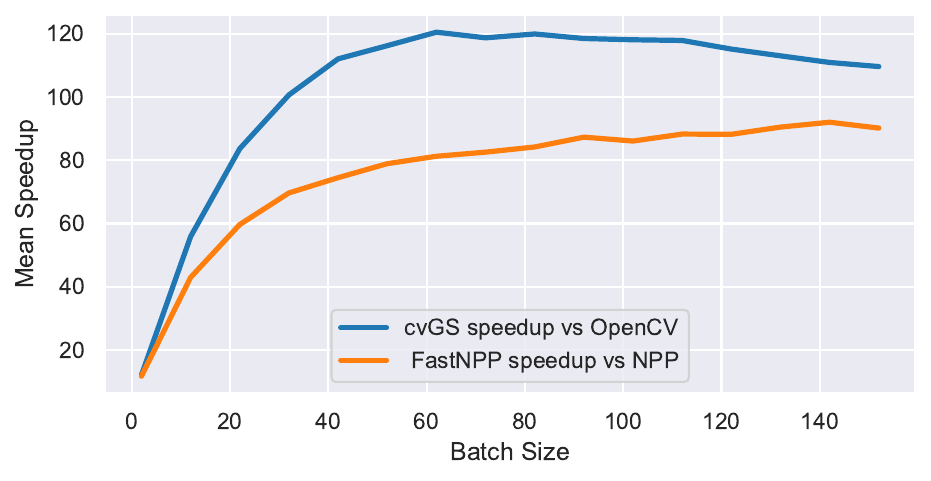}
		\caption{Speedup of the CPU part, when computing GPU parameters and launching kernels.} 
		\label{fig:CPUSpeedup}
	\end{figure}

        In our experiments we find that there is more speedup in the case of OpenCV-CUDA than \glsxtrshort{npp} (see in Figure~\ref{fig:CPUSpeedup}). This is probably due to a more efficient CPU code implementation in the \glsxtrshort{npp} functions. Since \glsxtrshort{npp} is closed source, unfortunately we cannot verify this assumption, but we note that in the OpenCV-CUDA version there are less kernel launches, and therefore less CUDA runtime related overhead.
        
    \subsection{Data Size}    
	In this experiment, we launch a kernel that processes a 1D number of \texttt{float} data elements. We set that number from 100 to \numprint{16654030} and evaluate how the amount of data to process affects the execution times of OpenCV-CUDA and cvGS.

        To avoid results being distorted by \glsxtrshort{hf} effects, we use a pipeline of \glsxtrshortpl{op} without batch. Specifically, we arrange a sequence of 100 pairs of \glsxtrshortpl{op}, where the first \glsxtrshort{op} is multiplication and the second is addition.
        The kernel configuration is $32\times 8\times 1$ threads per thread block and $N\times M\times 1$ thread blocks in the grid, where $N = width/32$ and $M = height/8$.
        
\begin{figure}
		\centering		\includegraphics[keepaspectratio=true,width=0.485\textwidth]{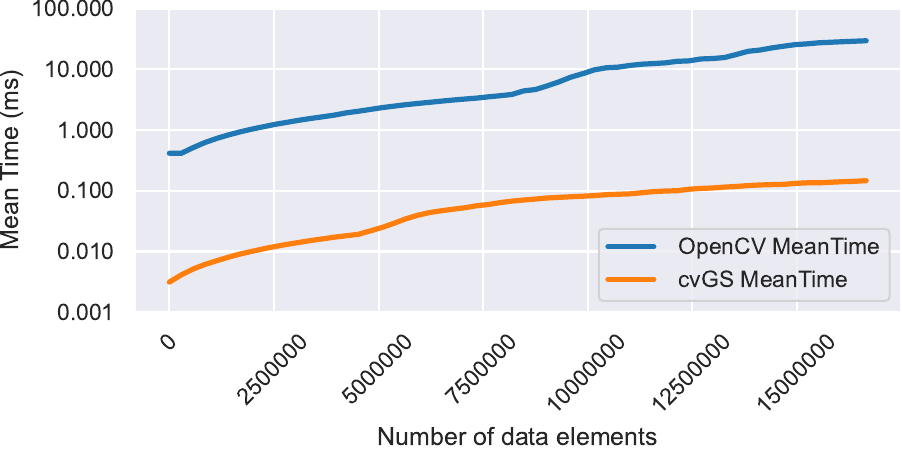}
		\caption{Execution times in logarithmic scale of OpenCV-CUDA and cvGS, according to the number of elements being processed.}
		\label{fig:DataSize}
	\end{figure}

        As we see in Figure~\ref{fig:DataSize}, the execution times of both OpenCV-CUDA and \glsxtrshort{cvgs} increase similarly, except for two interesting regions. In the first, close to zero, we see that the OpenCV-CUDA implementation, although slower than ours, does not increase its execution time for a range of data sizes from 100 to \numprint{282370}. In contrast, our implementation, although faster, increases the execution time on the same data size range. This is because the OpenCV-CUDA kernels are all \glsxtrshort{mb}, and the only factor limiting their execution time is GPU memory bandwidth.
        
        Inspecting the memory bandwidth utilization for those data sizes with NSight Compute, we observe that for 100 data elements, the GPU memory utilization is 0.6\%, whereas for \numprint{282370} elements it is 30\%. At about 40\% memory bandwidth utilization, the amount of data processed starts affecting the execution times of those kernels. In the case of the  \glsxtrshort{cvgs} fused kernel, we have observed in NSight Systems that the kernel is still \glsxtrshort{mb}, but very close to be balanced, so it is close to being \glsxtrshort{cb}. Therefore, the amount of elements to be processed affects the kernel execution time from the beginning.

        The second region of interest in Figure~\ref{fig:DataSize} is in the data size range of \numprint{9032740} to \numprint{16654030} elements. In that region, the execution time of the cvGS version grows more slowly while the OpenCV-CUDA version starts increasing faster. Profiling with NSight Compute, we explained this situation: at those data sizes, the GPU is approaching the memory bandwidth limit, attaining up to 90\%. This leads to memory access serialization and the warp stall reason being long score board (i.e., waiting for DRAM memory accesses), starts growing fast. In that situation, our kernel leverages considerably further compute instructions in order to exploit the GPU latency hiding than the markedly \glsxtrshort{mb} kernels in the OpenCV-CUDA version.
    
	\subsection{GPU Size}
        As we can see in Table~\ref{tab:systems}, different \glsxtrshortpl{gpu} feature different ratios of FP32 computing capacity versus GPU memory bandwidth. Following the definitions of \glsxtrshort{vf} and  \glsxtrshort{hf}, we would expect that larger \glsxtrshort{flopb} imply higher speedup when using \glsxtrshort{vf}. Therefore, a larger number of CUDA cores and memory bandwidth would imply higher speedup when using \glsxtrshort{hf}.

        In this experiment, we use the benchmark that has provided the greatest speedup (Section~\ref{sec:vfplushf}), to evaluate the benefit of \glsxtrshort{vf} (with fixed \glsxtrshort{hf} to batch 50) over \glsxtrshortpl{gpu} with different \glsxtrshort{flopb} proportions as seen in Table~\ref{tab:systems}. The kernel configuration is $32 \times 8 \times 1$ threads per thread block and $2\times 15\times 50$ threads in the grid.

\begin{figure}
		\centering
		\includegraphics[keepaspectratio=true,width=0.49\textwidth]{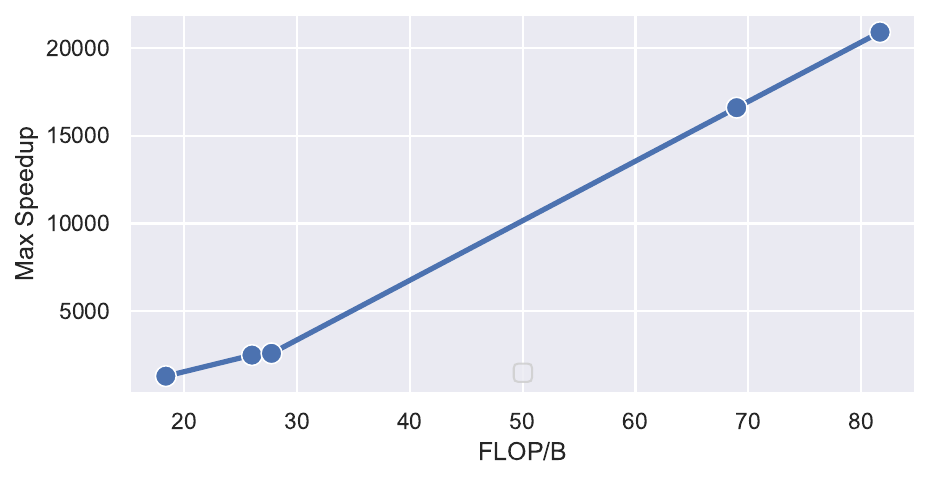}        
		\caption {Correlation of the maximum speedup found on each GPU when using  \glsxtrshort{vf} and \glsxtrshort{hf}, with the number of \glsxtrshort{flopb}. Each point in the chart corresponds to the GPU of one of the systems in Table~\ref{tab:systems}. The point with the lowest FLOP/B corresponds to S1, the next to S2 and so on until the point with the highest FLOP/B that corresponds to S5.} 
		\label{fig:GPUSize}
	\end{figure}

        We show the maximum speedup obtained on each system in fig \ref{fig:GPUSize}. As hypothesized, the maximum speedups are correlated with the number of \glsxtrshort{flopb}, attaining up to 20.9k$\times$ in System 5.

	\subsection{Data Type}
        In this experiment we evaluate the effects of the data type in the speedup between OpenCV-CUDA and \glsxtrshort{cvgs}. To that end, we fix a batch of 50 and $60\times120$ matrices. The operations performed are: $Batch(Read \rightarrow Cast \rightarrow Multiply \rightarrow Subtract \rightarrow Divide \rightarrow Write)$. We repeat this benchmark for a given combination of input and output types, in a set of 8 different combinations. The kernel configuration is $32 \times 8 \times 1$ threads per thread block and $2 \times 15\times 50$ thread blocks in the grid.
        
\begin{figure}
			\includegraphics[keepaspectratio=true,width=0.48\textwidth]{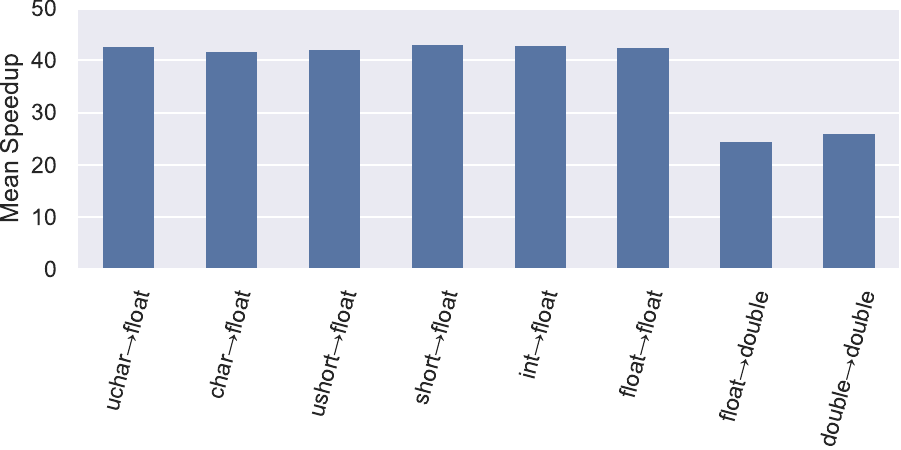}
		\caption{Speedup of \glsxtrshort{cvgs} versus OpenCV-CUDA, according to the input and output data types. In the chart, for each bar, the first type is the input data type, and the second after the arrow is the output data type.} 
		\label{fig:DataType}
	\end{figure}

        As we see in Figure~\ref{fig:DataType}, the speedup is similar for all types, except for those that compute with the {\tt double} data type. Using this data type in a GeForce GPU turns the kernel into \glsxtrshort{cb} easily, since every operation costs 64 more cycles than using float. Since \glsxtrshort{vf} is an optimization mostly for \glsxtrshort{mb} kernels, the more operations to perform, the more \glsxtrshort{cb} the kernel becomes and the smaller the speedup \glsxtrshort{vf} provides.

        Thus, the speedup increase in {\tt double}$\rightarrow${\tt double} versus {\tt float}$\rightarrow${\tt double} observed in Figure~\ref{fig:DataType} is explained by the fact that the {\tt double}$\rightarrow${\tt double} kernel is more \glsxtrshort{mb}. This is because both the {\tt float}$\rightarrow${\tt double} and {\tt double}$\rightarrow${\tt double} kernels perform double-precision computations. Since the amount of computation is the same and the {\tt double}$\rightarrow${\tt double} kernel has to read twice the amount of data, \glsxtrshort{vf} provides a greater speedup in the {\tt double}$\rightarrow${\tt double} kernel.

        \subsection{Comparing to NPP}
        We create a wrapper called FastNPP, with the functionality necessary for this test. The operations used are: $Batch(Crop \rightarrow Resize \rightarrow ColorConvert \rightarrow Multiply \rightarrow Subtract \rightarrow Divide \rightarrow Split)$. The resized images are $64\times 128$ pixels of \texttt{uchar3} type, stored in 3 planes of $64\times 128$ \texttt{float} each. Figure~\ref{fig:SyntaxComparisonNpp} illustrates the implementation in \glsxtrshort{npp} and its FastNPP counterpart.

        We do not compare with CUDA Graphs in this case, because using a syntax that is as similar to NPP as possible is part of the evaluation; hence, we compare performance at the same level of abstraction.

        In contrast to OpenCV-CUDA, \glsxtrshort{npp} features a version of resize that may execute a single kernel for a batch of images. We use that version to obtain the best possible performance. We vary the batch size from 10 to 150 and observe the speedup behavior. The kernel configuration is $32 \times 8 \times 1$ threads per thread block and $2 \times 16 \times N$ thread blocks in the grid, where $N$ varies from 10 to 150.

        As seen in Figure~\ref{fig:CPUSpeedup}, the CPU part of NPP overperforms OpenCV-CUDA. In contrast to both NPP and OpenCV-CUDA, our methodology enables to compute the CPU part of each \glsxtrshort{op} once and iteratively call the kernel with the same parameters. In this experiment we have performed two executions, one where with FastNPP we execute the CPU part on every iteration (as NPP does) and another where we execute the CPU part of the FastNPP version only once at the beginning. The goal is to assess the performance improvements of FastNPP in the two scenarios.
    
\begin{figure}
			\includegraphics[keepaspectratio=true,width=0.49\textwidth]{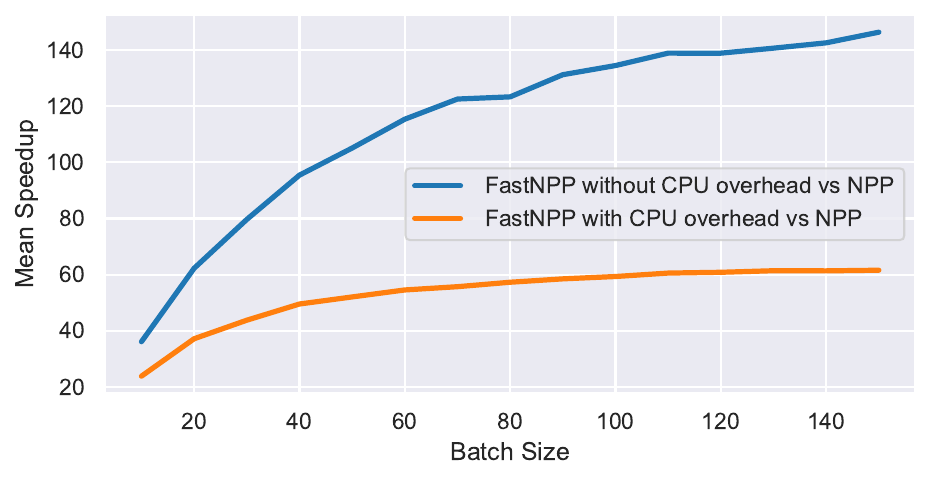}
		\caption{Speedup of FastNPP over NPP, with CPU precomputation in orange and without it in blue.} 
		\label{fig:test_npp_batchresize_x_split3D}
	\end{figure}

        In Figure~\ref{fig:test_npp_batchresize_x_split3D}, considering the CPU overhead, we see how the speedup increases with the batch size, but stagnates very rapidly at $61\times$. Instead, when precomputing the \glsxtrshortpl{iop}, the speedup starts already at $36\times$ and reaches $136\times$.

    \subsection{Library API Syntax Comparison} \label{exp_api}
\begin{figure}
    \centering
        \begin{subfigure}{0.5\textwidth} \includegraphics[keepaspectratio=true,width=1.0\textwidth]{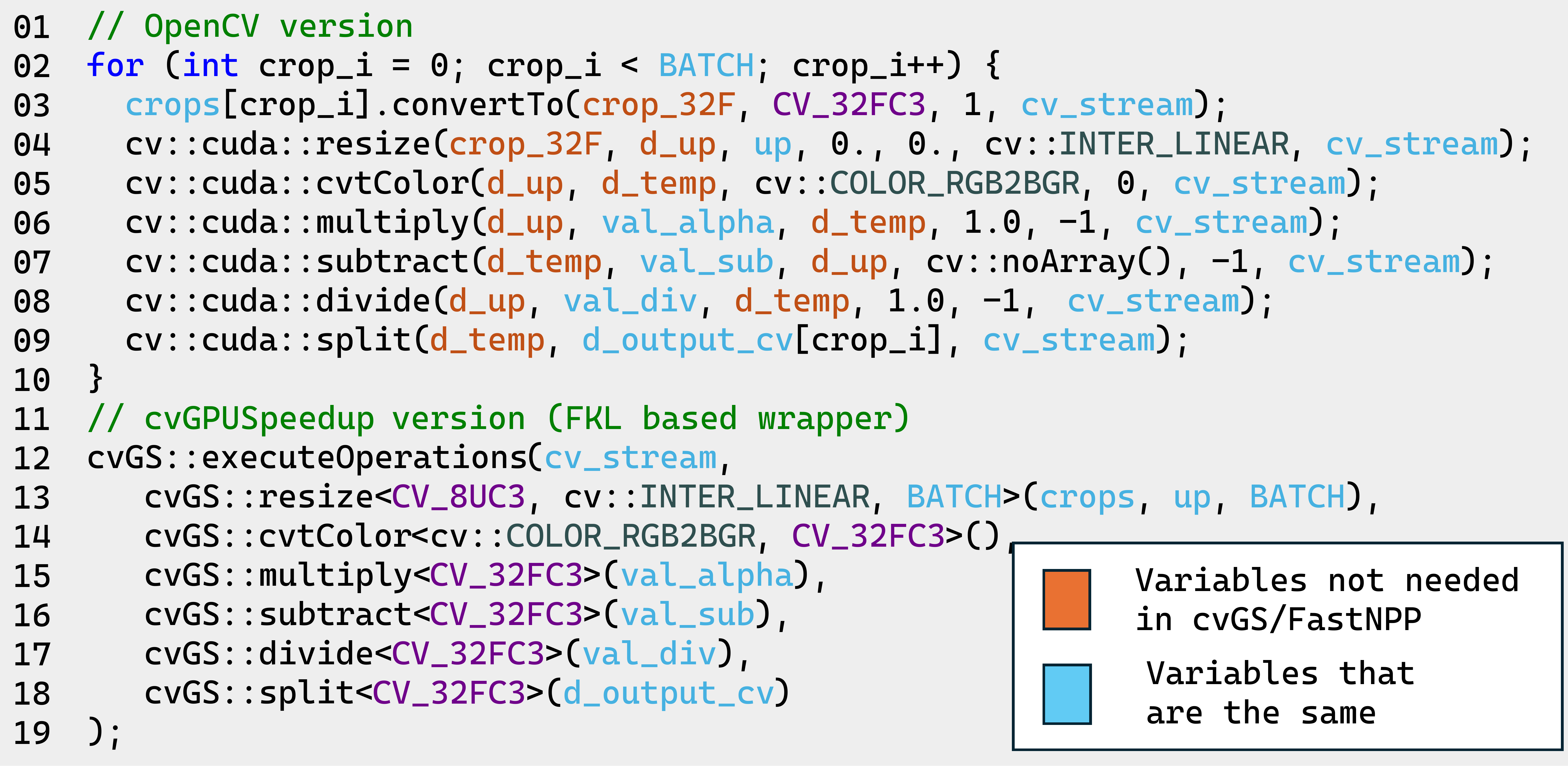}
            \caption{OpenCV-CUDA versus cvGS}		
            \label{fig:SyntaxComparisonOpenCV}
        \end{subfigure}
        \begin{subfigure}{0.5\textwidth} \includegraphics[keepaspectratio=true,width=1.0\textwidth]{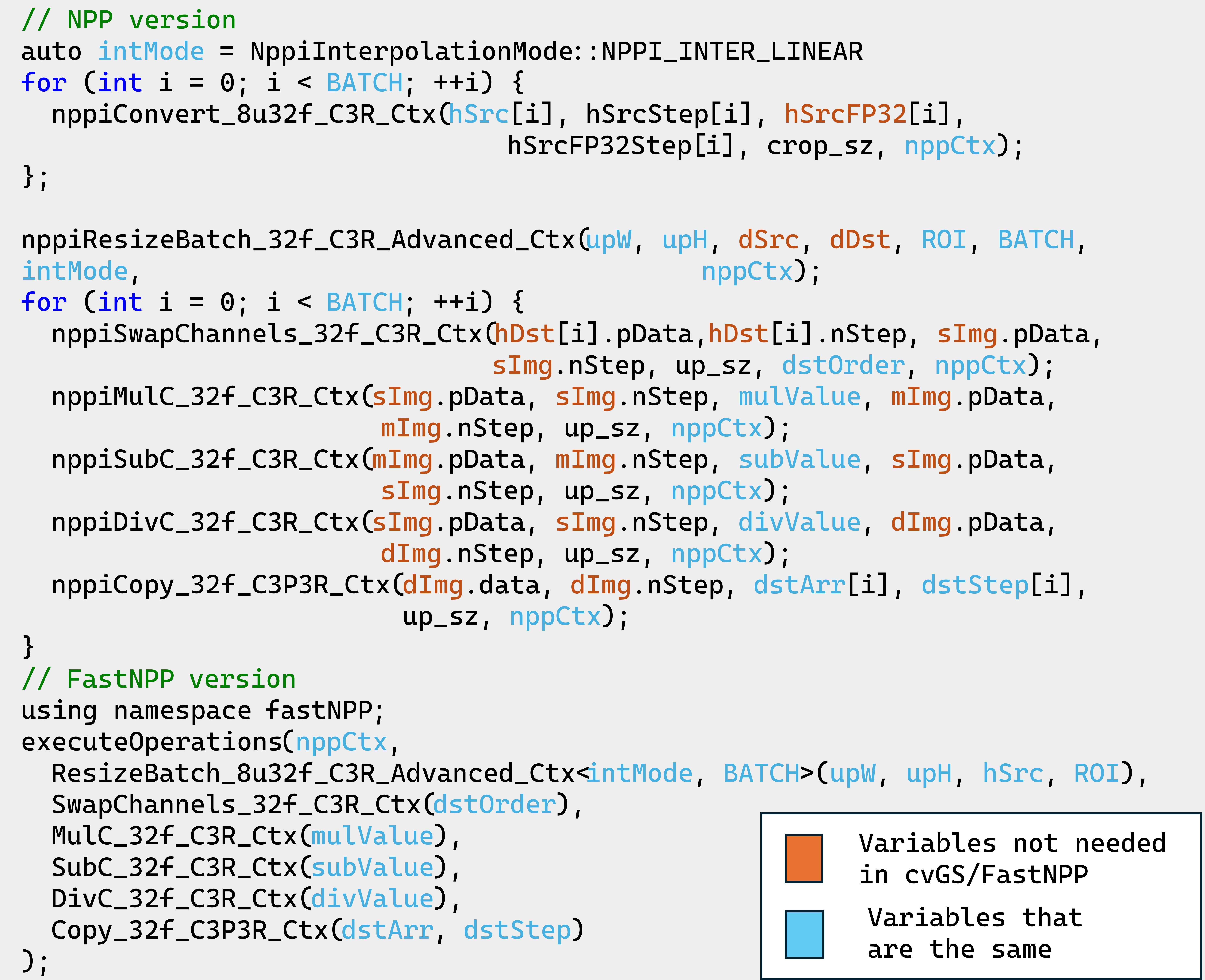}
            \caption{\glsxtrshort{npp} versus FastNPP}
            \label{fig:SyntaxComparisonNpp}
        \end{subfigure}
    \caption{Code example of a batched chain of operations, $Batch(Crop \rightarrow Resize \rightarrow ColorConvert \rightarrow Multiply \rightarrow Subtract \rightarrow Divide \rightarrow Split)$.}
    \label{fig:SyntaxComparison}
    \end{figure}

        In this section we show two examples of syntax, where we compare OpenCV-CUDA versus \glsxtrshort{cvgs} and \glsxtrshort{npp} versus FastNPP.
        As seen in Figure~\ref{fig:SyntaxComparison}, our two wrappers offer a similar and familiar syntax to the original libraries, yielding considerable speedup and saving GPU memory. The memory savings stem from the fact that we do not need to allocate several of the intermediate GPU pointers \glsxtrshort{npp} and OpenCV-CUDA require.

        In Figure~\ref{fig:SyntaxComparisonOpenCV} we have added a few template parameters, which help selecting code at compile time instead of runtime. In this case, final users often write these as literals; hence this does not pose additional difficulty for our target users. In the case of FastNPP, since the function name is already specifying the types, we do not need to add any template parameter. Future work may include solutions to reduce the number of template parameters requested to the final user, especially for the cvGS case.

        \subsection{GPU Memory Savings}
        We have computed the amount of memory saved by using our methodology. As shown in Figure~\ref{fig:SyntaxComparison},
        thanks to \glsxtrshort{vf}, we may avoid the allocation of several of the GPU memory regions that store intermediate images. Specifically, we avoid allocating the variables in orange \texttt{crop\_32F}, \texttt{d\_up} and \texttt{d\_temp}.

        Considering that each crop contains pixels in \texttt{float3} format, and that there are $60\times 120$ pixels per crop and \texttt{float3} = 12 bytes, we are saving 259KB of GPU memory in total. If we where working with larger images (such as 4k or 8k resolutions), the memory savings would be amplified. A 4k image in NV12 format uses 12.44MB of memory and in RGB it requires 24.88~MB, whereas an 8k image multiplies these sizes by a factor of 4.
    
\section{Conclusions and Future Work}
    In this article we have introduced novel methodology to express \glsxtrshort{hf} and \glsxtrshort{vf} leveraging standard C++, based on the concepts of \glsxtrshortpl{op}, \glsxtrshortpl{iop} and \glsxtrshortpl{dpp}.

    Our results show that, leveraging our methodology, we may easily imitate existing high-level library APIs, while performing automatic \glsxtrshort{hf} and \glsxtrshort{vf}. We have also demonstrated how these techniques produce large speedups, especially when combined, and more prominently when executing on large \glsxtrshortpl{gpu}. Additionally, we have also shown that the same technique provides GPU memory savings that may be very relevant for small systems with limited memory, or use cases where the input data for the kernel is very large. We provide the open-source implementations of \glsxtrshort{cvgs} \cite{cvgsrepo2025}, FastNPP \cite{fastnpprepo2025}, and the \glsxtrshort{fkl} \cite{fklrepo2025}, with which we performed the experiments.

    We have focused on demonstrating automatic \glsxtrshort{hf} and \glsxtrshort{vf} with a basic transform  \glsxtrshort{dpp} implementation, which already demonstrates a wide range of use cases and considerably large speedups.
    
	With our methodology, we open the door for the same \glsxtrshort{dpp} to be implemented on top of different hardware architectures such as multi-threaded CPUs, non-NVIDIA GPUs with different hardware features, NPUs, and any programmable architecture where its data parallelism may be expressed in C++17.

    Future work may include exploring C++20 Concepts to implement our definition of \glsxtrshortpl{op} and \glsxtrshortpl{iop}. Also, together with C++20 restrictions and other compile time utilities, these shall simplify both the development and debugging of our implementations.
    We also intend to explore those cases where users need to work with closed-source \glsxtrshortpl{dpp} and \glsxtrshortpl{op}, which shall provide beneficial results by using nvcc link-time optimizations.

    \section*{Acknowledgements}
    Authors thank Grup Mediapro for providing O. Amoros with time to work on this research and for using the FKL implementation in AutomaticTV in production.
    A. J. Peña is partially supported by the Ramón y Cajal fellowship RYC2020-030054-I funded by MCIN/AEI/ 10.13039/501100011033 and by ``ESF Investing in your future''.
	
    \section*{References}
    \let\oldsection=\section
    \renewcommand{\section}[2]{}%
    \printbibliography
     \let\section=\oldsection

    \section{Biography}
        \vskip -2\baselineskip plus -1fil
	\begin{IEEEbiographynophoto}{Oscar Amoros} is the CUDA Edge Computing Tech Lead at AutomaticTV (Grup Mediapro) and a PhD candidate in Computer Architecture at Universitat Politecnica de Catalunya. He is passionate about software performance, parallel computing and providing high level abstractions to hardware unaware programmers, that need to focus on solving their domain specific problems.
	\end{IEEEbiographynophoto}	
        \vskip -2\baselineskip plus -1fil
	\begin{IEEEbiographynophoto}{Albert Andaluz} is an independent researcher, who worked for more than a decade in the broadcast industry. Previously he was  research assistant at the \textit{interactive and augmented modeling group}(2008-2013) at the computer vision center, Autonomous University of Barcelona , under supervision of Prof. Debora Gil.  
    Albert holds a Master in Computer Vision and Artificial Intelligence (2009) from the Autonomous University of Barcelona , Spain.
	\end{IEEEbiographynophoto}
        \vskip -2\baselineskip plus -1fil
	\begin{IEEEbiographynophoto}{Johnny Nuñez}
        is a Developer Advocate for AI at NVIDIA. He is passionate about CUDA development and AI and has contributed to adapting numerous AI tools for edge devices. Johnny is also a member of the Jetson Research Lab, where he drives the development of AI and robotics on edge devices.
	\end{IEEEbiographynophoto}
        \vskip -2\baselineskip plus -1fil
	\begin{IEEEbiographynophoto}{Antonio J. Peña}
		is a Leading Researcher and Group Manager, leading the \textit{Accelerators and Communications for HPC} group at the Barcelona Supercomputing Center (BSC). His research interests in the area of runtime systems and programming models for high-performance computing include resource heterogeneity and communications.	
		Antonio holds a PhD in Advanced Computer Systems (2013), from Universitat Jaume I of Castelló, Spain.
	\end{IEEEbiographynophoto}	
\end{document}